\documentclass[prb,twocolumn,amsmath,amssymb,showpacs,nofootinbib,floatfix]{revtex4}

\usepackage{graphicx,bm}

\makeatletter
\def\graphicscale{\twocolumn@sw{0.3}{0.4}}
\def\graphicthreescale{\twocolumn@sw{0.3}{0.4}}

\begin{document}

\title{Universal behavior of two-dimensional bosonic gases \\
   at Berezinskii-Kosterlitz-Thouless transitions}

\author{Giacomo Ceccarelli$^1$, Jacopo Nespolo$^1$, 
Andrea Pelissetto$^2$, and Ettore Vicari$^1$} 

\address{$^1$ Dipartimento di Fisica dell'Universit\`a di Pisa
        and INFN, Largo Pontecorvo 3, I-56127 Pisa, Italy}
\address{$^2$ Dipartimento di Fisica dell'Universit\`a di Roma ``La Sapienza"
        and INFN, Sezione di Roma I, I-00185 Roma, Italy}

\date{May 13, 2013}

\begin{abstract}

We study the universal critical behavior of two-dimensional (2D)
lattice bosonic gases at the Berezinskii-Kosterlitz-Thouless (BKT)
transition, which separates the low-temperature superfluid phase from
the high-temperature normal phase. For this purpose, we perform
quantum Monte Carlo simulations of the hard-core Bose-Hubbard (BH)
model at zero chemical potential.  We determine the critical
temperature by using a matching method that relates finite-size data
for the BH model with corresponding data computed in the classical XY
model.  In this approach, the neglected scaling corrections decay as
inverse powers of the lattice size $L$, and not as powers of $1/\ln
L$, as in more standard approaches, making the estimate of the
critical temperature much more reliable.  Then, we consider the BH
model in the presence of a trapping harmonic potential, and verify the
universality of the trap-size dependence at the BKT critical point.
This issue is relevant for experiments with quasi-2D trapped cold
atoms.

\end{abstract}







\maketitle


\section{Introduction}
\label{intro}

Finite-temperature transitions in two-dimensional (2D) systems with a
global U(1) symmetry belong to the so-called 2D XY universality class,
and are described by the Berezinskii-Kosterlitz-Thouless (BKT)
theory.\cite{KT-73,B-72,Kosterlitz-74,JKKN-77} The low-temperature
phase is characterized by quasi-long range order (QLRO): Correlations
decay algebraically at large distances, without the emergence of a
nonvanishing order parameter.\cite{MW-66,H-67} If the temperature $T$
is above the BKT transition point $T_c$, these systems show instead a
standard disordered phase, with an exponential increase of the
correlation length $\xi$ associated with the critical modes as $T$
approaches $T_c$: $\xi\sim \exp(c /\sqrt{\tau})$, where $\tau\equiv
T/T_c-1$.

\begin{figure}[tbp]
\includegraphics*[scale=\graphicscale]{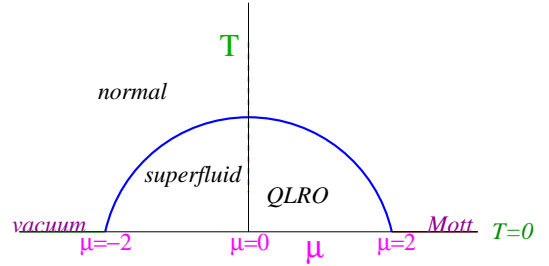}
\caption{(Color online) Sketch of the phase diagram of the 2D
  hard-core BH model or equivalently of the 2D XX spin model.  The
  finite-temperature BKT transition line separates the normal and superfluid
  QLRO phases.  It is symmetric with respect to $\mu\to -\mu$, and
  connects two $T=0$ quantum critical points at $\mu=\pm 2$.  The
  $T=0$ quantum transition at $\mu=-2$ separates the vacuum state and
  the superfluid phase, while the one at $\mu=2$ separates the
  superfluid phase from a $n=1$ Mott phase. }
\label{phd}
\end{figure}

Experimental evidences of BKT transitions have been reported for thin
films of liquid helium,\cite{BR-78,GKMD-08} superconducting
Josephson-junction arrays,\cite{RGBSN-81} quasi-2D trapped atomic
gases,\cite{HKCBD-06,KHD-07,HKCRD-08,CRRHP-09,HZGC-10,Pl-etal-11,Desb-etal-12}
including atomic systems constrained in quasi-2D optical
lattices.\cite{BDZ-08} Bosonic-atom systems in optical lattices are
effectively described\cite{JBCGZ-98} by the Bose-Hubbard (BH) model
\cite{FWGF-89}
\begin{eqnarray}
H_{\rm BH} &=& - {J\over 2} \sum_{\langle {\bf x}{\bf y}\rangle} (b_{\bf x}^\dagger b_{\bf y}+
b_{\bf y}^\dagger b_{\bf x}) 
\label{bhm}\\
&&+ {U\over 2} \sum_{\bf x} n_{\bf x}(n_{\bf x}-1) - \mu \sum_{\bf x} n_{\bf x},\nonumber
\end{eqnarray}
where $b_{\bf x}$ is a bosonic operator, $n_{\bf x}\equiv b_{\bf
  x}^\dagger b_{\bf x}$ is the particle density operator, and the sums
run over the bonds ${\langle {\bf x} {\bf y} \rangle }$ and the sites
${\bf x}$ of a square lattice.  The phase diagram of the 2D BH model
presents finite-temperature BKT transition lines separating the normal
phase from the superfluid QLRO phase.  In the hard-core limit $U\to
\infty$, the BH model can be exactly mapped onto the so-called XX spin
model:\cite{mappingXX}
\begin{eqnarray}
H_{\rm XX} = - J \sum_{\langle {\bf x}{\bf y}\rangle} 
\left( S^1_{\bf x} S^1_{\bf y} + 
S^2_{\bf x} S^2_{\bf y} \right) + \mu \sum_{\bf x} S^3_{\bf x},
\label{XX}
\end{eqnarray}
where $S^a_{\bf x}=\sigma^a_{\bf x}/2$ and $\sigma^a$ are the 
Pauli matrices.  
 A sketch of the phase diagram in the hard-core
limit is shown in Fig.~\ref{phd}.  

The BKT transition is characterized by logarithmic corrections to the
asymptotic behavior, due to the presence of marginal
renormalization-group (RG) perturbations at the BKT fixed
point.\cite{AGG-80,HMP-94,HP-97,Balog-01,Hasenbusch-05,PV-13} This makes the
numerical or experimental determination of the critical parameters
quite difficult. Indeed, logarithmic corrections cannot be easily
detected, and therefore taken into account, by comparing data in the
relative small range of parameters close to criticality, which are
available from experiments or numerical simulations.  Therefore, the
extrapolations of the numerical and experimental results, which must
be performed to determine the universal critical behavior, are often
not fully reliable and subject to systematic errors which are quite
difficult to estimate.

In this paper we report a numerical investigation of the universal
behavior of the 2D hard-core BH model at the BKT transition and in the
superfluid QLRO phase. For this purpose, we perform quantum Monte
Carlo (QMC) simulations at zero chemical potential on square lattices
$L^2$, up to $L=256$, with periodic boundary conditions. We verify the
spin-wave predictions for the behavior in the QLRO phase.  Then we
present an accurate determination of the BKT transition temperature,
using a matching method, which generalizes the approach of Hasenbusch,
Marcu, and
Pinn.\cite{HMP-94,HP-97,Hasenbusch-05,Hasenbusch-08,Hasenbusch-09,Hasenbusch-12}
The critical temperature is obtained by matching the finite-size
scaling (FSS) behavior of RG invariant quantities computed in the BH
model with the FSS behavior of the same quantities computed in the
classical 2D XY model,
\begin{equation}
{\cal H}_{\rm XY} = - J \sum_{\langle {\bf x}{\bf y} \rangle } {\rm
  Re} \,  \bar{\psi}_{\bf x} \psi_{\bf y}, \qquad \psi_{\bf x} \in {\rm U}(1),
\label{XYmodel}
\end{equation}
for which the critical temperature is known with high
accuracy:\cite{HP-97,Hasenbusch-05,KO-12} $\beta_{\rm XY} \equiv
1/T_{\rm XY}=1.1199(1)$ setting $J=1$ (see, however, Ref.~\onlinecite{HKS-13}
for a different estimate).  In this method the neglected scaling
corrections are of order $L^{-\epsilon}$, $\epsilon\gtrsim 2$ for our
choices of variables. This is a crucial improvement with respect to
more standard approaches, in which the neglected corrections decay as
powers of $1/\ln L$.

Experiments with cold atoms~\cite{CWK-02,BDZ-08} are usually performed
in the presence of a trapping potential, which can be taken into
account by adding a corresponding term in the Hamiltonian of the BH
model,
\begin{eqnarray}
&& H_{\rm tBH} =  H_{\rm BH} + \sum_{\bf x} V(r_{\bf x}) n_{\bf x}, 
\label{bhmt}\\
&& V(r)= u^p r^p,\label{potential}
\end{eqnarray}
where $r$ is the distance from the center of the trap, and $p$ is a
positive even exponent.  The trap size is defined by
\begin{equation}
\ell_t \equiv J^{1/p}/u.  
\label{trapsize}
\end{equation}
The trapping potential is effectively harmonic in most experiments,
i.e. $p=2$.  The BKT critical behavior is significantly modified by
the presence of the trap.\cite{CR-12,HK-08}  An accurate
experimental determination of the critical parameters, such as the
critical temperature, critical exponents, etc..., in trapped-particle
systems requires a quantitative analysis of the trap effects.

The inhomogeneity due to the trapping potential drastically changes,
even qualitatively, the general features of the critical behavior.
For example, the correlation functions of the critical modes do not
develop a diverging length scale in a trap.  Nevertheless, when the
trap size $\ell_t$ becomes large, the system develops a critical
scaling behavior, which can be described in the framework of the
trap-size scaling (TSS) theory.\cite{CV-09,CV-10} TSS has some
analogies with the standard FSS for homogeneous systems with two main
differences. First, the system is inhomogeneous, due to the
space-dependence of the external field.  Second, at the critical
point, the critical correlation length $\xi_t$ is not simply
proportional to the trap size $\ell_t$ generally, but satisfies a
nontrivial scaling relation $\xi_t\sim \ell_t^\theta$, where $\theta$
is the trap exponent.  The analysis of the RG flow at a BKT
transition~\cite{PV-13} shows that the TSS of the 2D BH model
(\ref{bhmt}) is characterized by the trap exponent $\theta=1$, with
additional multiplicative logarithms: for example, at the BKT critical
point $\xi_t \sim \ell_t (\ln \ell_t)^{-\kappa}$, where the exponent
$\kappa$ depends on the general features of the trap.

In this paper we also present a numerical QMC study of the 2D
hard-core BH model in the presence of a harmonic potential. We compare
the QMC data with the theoretical predictions obtained from the
analysis of the BKT RG flow in the presence of a trap.  We argue, and
provide numerical evidence, that the critical trap-size dependence for
$\mu\le 0$ shares universal features with the inhomogeneous 2D
classical XY model~\cite{PV-13,CV-11}
\begin{eqnarray}
&& {\cal H}_U = -  J \sum_{\langle {\bf x}{\bf y} \rangle }
 {\rm Re} \, \bar{\psi}_{\bf x} U_{{\bf x}{\bf y}} \psi_{\bf y},
\label{xymodtr} \\
&& U_{{\bf x}{\bf y}} = [1 + W(r_{{\bf x}{\bf y}})]^{-1}, \quad W(r)=u^q r^q,
\label{hoppingt}
\end{eqnarray}
where $q$ is an even positive integer, $r_{{\bf x}{\bf y}}$ is the
distance from the origin of the midpoint of the lattice link
connecting the nearest-neighbor sites ${\bf x}$ and ${\bf y}$.  This
U(1)-symmetric model may be considered as a classical XY model with an
effective space-dependent temperature
\begin{equation}
T_{\rm eff}(T,{\bf x})=T\,[1+W(r)]\ge T.  
\label{teffxy}
\end{equation}
A length scale analogous to the trap size can be defined as
$\ell_t\sim 1/u$. We will show that the correspondence $p=q$ holds for
$\mu<0$, while $p=2q$ at half filling $\mu=0$.

The paper is organized as follows.  In Sec.~\ref{sec2} we present a
FSS analysis of QMC data of the 2D hard-core BH model along the
$\mu=0$ line of the phase diagram presented in Fig.~\ref{phd}.  We
show how the critical temperature can be accurately determined by a
matching method, that relates the finite-size behavior of
dimensionless RG invariant quantities in the BH model with that of the
same quantities in the classical XY model. In this approach the
neglected scaling corrections decay as inverse powers of the lattice
size, making the method much more robust and accurate with respect to
standard approaches, affected by logarithmic corrections.  Method and
results are presented in detail, to provide a thorough account of the
accuracy and the reliability of the estimate of the critical
temperature we obtain.  In Sec.~\ref{trapan} we investigate the
universal critical behavior of the BH model in the presence of a
trapping potential.  Finally, in Sec.~\ref{conclusions} we summarize
our main results and draw our conclusions.  Technical details are
reported in the appendices.

\section{Finite-size scaling of the 2D BH model}
\label{sec2}

We consider the $U\to\infty$ hard-core limit of the BH model at zero
chemical potential $\mu=0$, and study the critical behavior at the
BKT transition and in the low-temperature QLRO phase.  In the hard-core limit,
the $\mu=0$ line corresponds to half filling, i.e.,
\begin{equation}
\rho = \langle n_{\bf x} \rangle = \langle b_{\bf x}^\dagger b_{\bf x}\rangle = 1/2
\label{rhoh}
\end{equation}
for any $T$. In the XX model (\ref{XX}), this condition corresponds to
$\langle S_{\bf x}^3 \rangle = 0$, a result which can be inferred by
using the $Z_2$-symmetry of the Hamiltonian in the absence of an
external magnetic field.

We consider homogeneous $L\times L$ square systems with periodic
boundary conditions.  We present a finite-size scaling (FSS) analysis
of QMC data up to $L=256$.  The QMC simulations are performed using
the stochastic series expansion algorithm with the directed
operator-loop technique.\cite{Sandvik-99,SS-02,DT-01,stats}

\subsection{Observables}
\label{sec21}

We compute the one-particle correlation function
\begin{equation}
G({\bf x},{\bf y}) \equiv \langle b_{\bf x}^\dagger  b_{\bf y} \rangle.
\label{gbdef}
\end{equation}
Due to translation invariance, it only depends on the difference of
the arguments, i.e.,  $G({\bf x},{\bf y}) \equiv G({\bf x}-{\bf y})$.
We consider the susceptibility
\begin{equation}
\chi =  \sum_{{\bf x}} G({\bf x}),
\label{chisusc}
\end{equation}
which is the zero-momentum component of the Fourier transform
$\widetilde{G}({\bf k}) = \sum_{{\bf x}} e^{i{\bf k}\cdot{\bf x}}
G({\bf x})$.  The second-moment correlation length $\xi$ is
conveniently defined by
\begin{equation}
\xi^2 \equiv  {1\over 4 \sin^2 (p_{\rm min}/2)} 
{\widetilde{G}({\bf 0}) - \widetilde{G}({\bf p})\over 
\widetilde{G}({\bf p})},
\label{xidefpb}
\end{equation}
where ${\bf p} = (p_{\rm min},0)$, $p_{\rm min} \equiv 2 \pi/L$.

In the studies of critical phenomena, dimensionless RG invariant
quantities $R$ are particularly useful to determine the critical
parameters.  Beside the ratio
\begin{equation}
X \equiv \xi/L,
\label{Xdef}
\end{equation}
we consider the so-called helicity modulus~\cite{CHPV-06,footnoteY}
$\Upsilon$, which is related to the spin stiffness in spin
models,\cite{Sandvik-97} and to the superfluid density in particle
systems.\cite{FBJ-73,PC-87} In QMC simulations using the stochastic
series expansion algorithm, $\Upsilon$ is obtained from the linear
winding number $w_i$ along the $i^{\rm th}$
direction,\cite{Sandvik-97}
\begin{equation}
\Upsilon =  \langle w_i^2 \rangle,
\qquad w_i = \frac{N_i^+ - N_i^-}{L},
\label{ulw}
\end{equation}
where $N_i^+$ and $N_i^-$ are the number of non-diagonal operators
which move the particles in the positive and negative
$i^{\rm th}$ direction, respectively. 

\subsection{RG analysis of the universal FSS} \label{sec22}

The analysis of the RG flow at the BKT
transition \cite{AGG-80,HMP-94,HP-97,Balog-01,Hasenbusch-05,PV-13} allows us 
to express any RG invariant quantity $R$ in terms of two 
(universally defined) nonlinear scaling fields $v(L/\Lambda,Q)$ and $Q$, 
see Ref.~\onlinecite{PV-13} and Appendix \ref{AppA}. 
The nonuniversal details that characterize the 
model are encoded in the model-dependent scale $\Lambda$ and in the 
temperature dependence of $Q$,
\begin{equation}
Q = \alpha_1 (T-T_c) + \alpha_2 (T - T_c)^2 + \ldots
\label{Q-vsT}
\end{equation}
where $\alpha_1$, $\alpha_2$, $\ldots$ 
depend on the model. In terms of these two
quantities we can write the scaling expressions
\begin{equation}
R(L,T) = {\cal R}[v(L/\Lambda,Q),Q],
\label{RLT-RG}
\end{equation}
where ${\cal R}(x,y)$ depends on the shape of the system
(e.g. the aspect ratio of the lattice) and on the boundary conditions, 
but is independent of the microscopic details, i.e. it is universal.
At the critical point $T_c$, we have $Q=0$
and 
\begin{equation}
R(L,T_c) = {\cal R}[v(L/\Lambda,0),0] = {\cal R}_c(L/\Lambda).
\label{RLTC}
\end{equation}

Eq.~(\ref{RLTC}) can be used to obtain a matching relation that will 
be the basis of our numerical method to estimate the critical temperature.
Consider two different models with critical temperatures $T_{c1}$ and $T_{c2}$,
respectively, and let $R^{(1)}(L,T)$ and $R^{(2)}(L,T)$ be 
the estimates of the RG invariant quantity $R$ in the two models. 
Using Eq.~(\ref{RLTC}) we can write 
\begin{equation}
R^{(1)}(L,T_{c1}) ={\cal R}_c(L/\Lambda_1), \;\;
R^{(2)}(L,T_{c2}) ={\cal R}_c(L/\Lambda_2),
\end{equation}
with two different nonuniversal constants $\Lambda_1$ and $\Lambda_2$. 
If we now define $\lambda = \Lambda_1/\Lambda_2$, we obtain 
\begin{equation}
R^{(1)}(\lambda L,T_{c1}) = R^{(2)}(L,T_{c2}), 
\label{matching-R}
\end{equation}
which simply states that the size dependence of $R$ at the critical 
point is the same in the two models, as long as one considers values of $L$ 
that differ by a factor $\lambda$. Analogous relations hold for 
the temperature derivatives of $R$ computed at the critical point.
For instance, by using Eqs.~(\ref{RLT-RG}) and (\ref{Q-vsT}) we can write
\begin{equation}
S(L) \equiv \left. {\partial R\over \partial T}\right|_{T_c} = 
   \alpha_1 {\cal S}_c(L/\Lambda),
\end{equation}
where ${\cal S}_c(L/\Lambda)$ is universal, all model dependence being 
included in the length scale $\Lambda$ and in the constant $\alpha_1$. 
This expression implies the matching condition
\begin{equation}
S^{(1)}(\lambda L) = k S^{(2)}(L),
\label{matching-S}
\end{equation}
where $k = \alpha_1^{(1)}/\alpha_1^{(2)}$ is the ratio of the nonuniversal
constants $\alpha_1$ in the two models.

It is important to stress that relation (\ref{RLT-RG}), and therefore 
also Eqs.~(\ref{matching-R}) and (\ref{matching-S}), are obtained by 
considering only the marginal operators that characterize the BKT 
transitions. There are several sources of corrections to these 
relations. First, one should consider the subleading irrelevant operators,
which give rise to scaling corrections that decay as inverse powers of $L$.
According to the standard spin-wave theory, the most relevant one 
has RG dimension $-2$, hence it gives rise to corrections of order 
$1/L^2$ (an additional logarithmic factor $\ln^p L$ can also be present, due to
possible resonances between the subleading and the marginal operators).
For the helicity modulus this is the only source of scaling corrections, 
hence 
\begin{eqnarray}
&& \Upsilon(L,T_c) = {\cal Y}_c(L/\Lambda) + O(L^{-2}), 
\label{yscal} \\
&& \Upsilon^{(1)}(\lambda L,T_{c1}) = \Upsilon^{(2)}(L,T_{c2}) + O(L^{-2}). 
  \label{matching-Y}
\end{eqnarray}
On the other hand, $\chi$ and $\xi$, and therefore $X=\xi/L$, get also
contributions from the analytic background at the
transition, \cite{PV-02} which gives rise to corrections
of order $1/L^{2-\eta} = L^{-7/4}$. Therefore, in the case of $X$ we expect
\begin{eqnarray}
X(L,T_c) = {\cal X}_c(L/\Lambda) + O(L^{-7/4}).\label{xscal}
\end{eqnarray}

\subsection{Spin-wave behavior}
\label{sec23}

\begin{figure}[tbp]
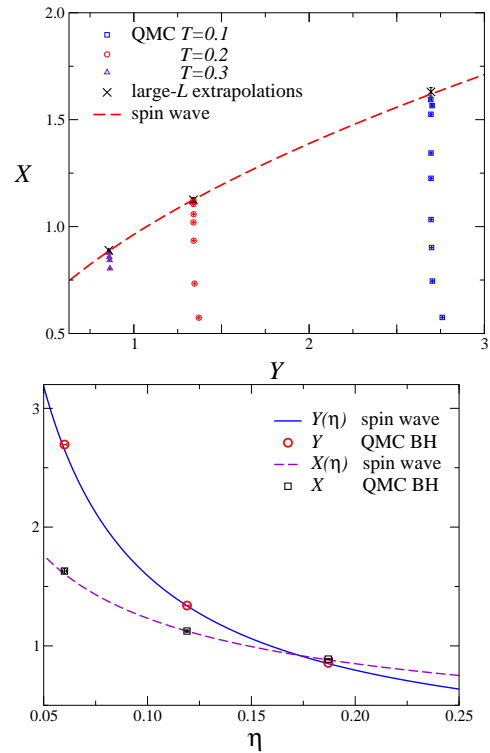

\includegraphics*[scale=\graphicscale]{fig2a.eps}
\includegraphics*[scale=\graphicscale]{fig2b.eps}
\caption{(Color online) Universal relations among the helicity modulus
  $\Upsilon$, the ratio $X\equiv\xi/L$, and the exponent $\eta$ for a square
  system with periodic boundary conditions in the low-temperature QLRO
  phase.  We report QMC data for the 2D hard-core BH model at
  $T=0.1,\,0.2,\,0.3$ up to $L=128$, and the spin-wave predictions. 
  Top panel:
  numerical estimates of $X$ versus $\Upsilon$, corresponding large-$L$ 
  extrapolations, and spin-wave prediction.
  Bottom panel: 
  large-$L$ extrapolations of the numerical results and
  spin-wave curves $X(\eta)$, $\Upsilon(\eta)$ versus $\eta$.
  }
\label{gqlro}
\end{figure}

The spin-wave theory describes the critical behavior of the model 
along the line of fixed points that runs from 
$T=0$ up to the BKT point $T_c$.  Conformal field
theory (CFT) exactly provides the large-$L$ limit of the one-particle
function in the spin-wave model, hence 
it allows one to compute the large-$L$ limit of $X$ and $\Upsilon$ 
and the exponent $\eta$ as a function of the spin-wave coupling, 
see App.~\ref{gauuniv} and Ref.~\onlinecite{Hasenbusch-05} for details. 
These relations depend on the nonuniversal spin-wave coupling, hence the 
results cannot be compared directly with those obtained in other models that 
have the same line of fixed points. Universal relations can be obtained by 
eliminating the spin-wave coupling,\cite{Hasenbusch-05} 
i.e., by expressing $X$ and $\Upsilon$
in terms of $\eta$, or $X$ in terms of $\Upsilon$. These relations do not 
depend on the model, but, as usual for FSS properties, they 
depend on the aspect ratio of the system and on the boundary conditions. 

In Fig.~\ref{gqlro} we show the curves obtained in the spin-wave
theory for a square lattice with periodic boundary conditions,
\cite{Hasenbusch-05} and compare them with the large-$L$
extrapolations of the QMC results at $T=0.1,\,0.2,\,0.3$ (in units of
$J$), which all belong to the QLRO phase.  They are obtained by
fitting the available QMC estimates of $\chi$, $X$, and $\Upsilon$ for
several values of $L$ up to $L=128$ to
\begin{eqnarray}
&&\ln\chi(L) = a + (2-\eta)\ln L + b L^{-\varepsilon}, \label{fitchilt}\\
&& X(L) = X + a L^{-\varepsilon}, \label{fitXlt}\\
&& \Upsilon(L) = \Upsilon + a L^{-\zeta}, \label{fitYlt}
\end{eqnarray}
respectively, where $\varepsilon$ and $\zeta$ are the exponents
associated with the expected leading corrections:\cite{PV-13,HPV-05}
\begin{eqnarray}
&&\varepsilon={\rm Min}[2-\eta,\omega],\qquad
\zeta={\rm Min}[2,\omega],\label{varezeta}\\
&&\omega=1/\eta-4 + O[(1/\eta-4)^2].
\label{omega}
\end{eqnarray}
We obtain $\eta=0.060(1),\,0.119(1),\,0.187(2)$ for $T=0.1,\,0.2,\,0.3$, 
respectively, suggesting 
\begin{equation}
\eta\approx 0.60 \,T 
\label{etat}
\end{equation}
for small values of $T$. The large-$L$ extrapolations of $X$ and
$\Upsilon$ are in agreement with the spin-wave predictions, as shown
in Fig.~\ref{gqlro} (the error on the large-$L$ extrapolations takes
into account the uncertainty on $\omega$).  Moreover, they also
confirm that $T_c>0.3$.

At $T_c$, the asymptotic large-$L$ behavior can be derived from the RG
prediction (\ref{RLT-RG}).  For $L\to\infty$ the scaling field $v$ can
be expanded as \cite{AGG-80,Balog-01,Hasenbusch-05,PV-13}
\begin{equation}
v(L/\Lambda,0) = {1\over w} + O \left({\ln w\over w^3}\right),
\end{equation}
with 
\begin{equation}
w = \ln {L\over \Lambda} + {1\over2} \ln\ln {L\over \Lambda}.
\label{def-upsilon}
\end{equation}
Expanding ${\cal R}(v,0)$ in powers of $v$, we obtain 
\begin{equation}
R(L,T_c) = R^* + C_R w^{-1} + O(w^{-2}).
\label{rasym}
\end{equation}
The constants $R^*$ and $C_R$ can be computed by using again spin-wave
theory, obtaining\cite{Hasenbusch-05} for $X$ and $\Upsilon$ the accurate 
numbers (see App.~\ref{gauuniv})
\begin{eqnarray}
&Y^* = 0.6365081789,\quad & C_Y = 0.31889945,
\label{retalo}\\
&X^* = 0.7506912222, \quad & C_X = 0.21243137.
\label{rxilo}
\end{eqnarray}

\subsection{Estimate of $T_c$ for the Bose-Hubbard model}
\label{sec24}

\subsubsection{The matching method} \label{sec24.1}

The presence of marginal RG perturbations at the BKT fixed point gives
rise to logarithmic corrections to the asymptotic scaling
behavior.\cite{AGG-80,HMP-94,HP-97,Balog-01,Hasenbusch-05,PV-13} In
FSS studies these slowly decaying corrections make an accurate
determination of the critical temperature quite difficult, because
they drastically affect the reliability of the large-$L$
extrapolations of the numerical results.  To overcome this problem, we
employ a matching method that generalizes the approach used in
Refs.~\onlinecite{HMP-94,HP-97} to obtain accurate estimates of the
critical temperature for the roughening transition of various
solid-on-solid models and the magnetic transition of the classical XY
model. Hasenbusch and Pinn\cite{HP-97} computed the RG flow both in
the XY model and in the body-centered solid-on-solid (BCSOS) model for
which $T_c$ is exactly known, and then determined $T_{c,XY}$ by
matching the results for the two models. Here we use a similar idea,
matching the finite-size behavior of RG invariant quantities.

To clarify the method, suppose there is a model belonging to the BKT
universality class for which the critical temperature $T_{c1}$ is
known. Moreover, assume that, for this model, a RG invariant quantity
$R$ is known precisely as a function of the size $L$, i.e., the
function $R^{(1)}(L,T_{c1})$ defined in Sec.~\ref{sec22} is known with
high precision for all values of $L$. Now, consider a second model,
whose critical temperature $T_{c2}$ is not known, and compute the same
RG invariant quantity for several values of $L$ and $T$, i.e., the
function $R^{(2)}(L,T)$. Then, $T_{c2}$ is determined by the matching
condition (\ref{matching-R}): The rescaling constant $\lambda$ and
$T_{c2}$ are obtained by minimizing the difference between
$R^{(1)}(\lambda L,T_{c1})$ and $R^{(2)}(L,T_{c2})$. Note that the
neglected scaling corrections in this two-parameter fit are of order
$L^{-\epsilon}$ ($\epsilon=2$ for $\Upsilon$ and $\epsilon = 7/4$ for
$X$), hence we overcome the problem of the logarithmic corrections.

In this work we take the classical XY model as reference model.  The
critical temperature is known quite precisely, \cite{HP-97}
$\beta_{c,\rm XY}\equiv 1/T_{c,\rm XY}=1.1199(1)$, though not
exactly. This introduces a systematic uncertainty in the procedure,
which can be accurately quantified, as we shall discuss below in
Sec.~\ref{sec24.4}.  We consider the RG invariant helicity modulus
$\Upsilon$ and ratio $X\equiv \xi/L$, to look for the optimal
matching.  For this purpose, we need accurate estimates of
$\widetilde{\Upsilon}_{\rm XY}(L)\equiv \Upsilon(L,T_{c,\rm XY})$ and
$\widetilde{X}_{\rm XY}(L) \equiv X(L,T_{c,\rm XY})$. They are
obtained by interpolating the available XY data (which extend up to
$L=4096$, see Refs.~\onlinecite{Hasenbusch-05,Hasenbusch-08}) and
taking into account the large-$L$ asymptotic expansions
(\ref{rasym}). The resulting expressions are reported in
App.~\ref{xycurves}. The XY results are compared with BH results
corresponding to lattice sizes $4\le L \le 256$ and temperatures in
the interval $0.3415 \le T \le 0.3450$, which is the interval of
temperatures in which $T_c$ is expected to lie, according to previous
works~\cite{CR-12,HK-97,Ding-92,DM-90}.

We would like to stress that the matching method is general and can be used
to determine the critical temperature  of any physical system undergoing
a BKT transition, using in all cases the interpolation curves reported in
App.~\ref{xycurves}.

\subsubsection{Direct analyses} \label{sec24.2}

\begin{figure}[tbp]
\includegraphics[width=8.5cm, keepaspectratio=true]{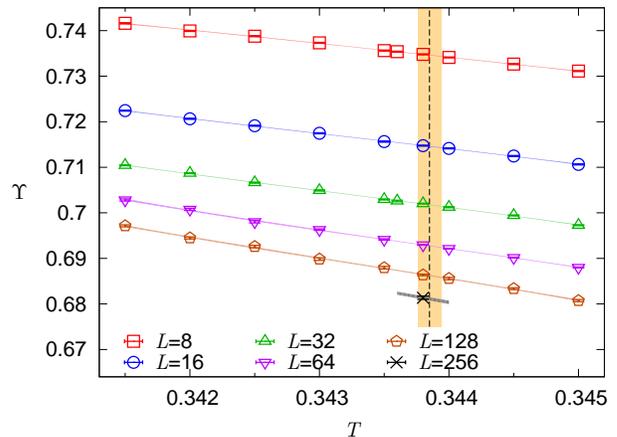}
\caption{(Color online) QMC estimates of $\Upsilon$ for several
  lattice sizes $L$
  up to $L=256$.  For $L\le 128$, the lines are interpolations of the data. 
  For $L=256$, $\Upsilon$ was only computed for $T=0.3438$: the reported line
  corresponds to an extrapolation that uses the first two terms of 
  the expansion of $\Upsilon$ around $T=0.3438$.
The vertical band shows the final estimate $T_c=0.34385(9)$. 
}
\label{rawY}
\end{figure}

\begin{table}
\caption{ Results for $\widetilde{T}(L)$ and $\lambda(L)$ 
  as obtained by solving
  Eq.~(\ref{twopm}).  We also report $T_c$ as
  obtained by linear fits to $\widetilde{T}(L)=T_c + c/L^2$ of the data
  satisfying $L\ge L_{\rm min}$, and the corresponding $\chi^2/{\rm dof}$,
  where dof is the number of degrees of freedom of the fit.
}
\label{restwopY}
\begin{ruledtabular}
\begin{tabular}{rllcrlc}
$L$ & $\widetilde{T}(L)$ & $\lambda(L)$ & $\;\;$ & $L_{\rm min}$& $T_c$ & 
$\chi^2/{\rm dof}$ \\
\colrule
4  & 0.34344(4)  & 1.457(3)   & & 4  & 0.34384(3) & 0.8 \\
6  & 0.34380(7)  & 1.499(8)   & & 6  & 0.34381(3) & 0.4 \\
8  & 0.34379(6)  & 1.501(9)   & & 8  & 0.34382(3) & 0.5 \\
12 & 0.34378(6)  & 1.50(1)  & & 12 & 0.34383(4) & 0.5 \\
16 & 0.34381(6)  & 1.51(2)  & & 16 & 0.34384(5) & 0.6 \\
24 & 0.34378(9)  & 1.50(3) & & 24 & 0.34389(7) & 0.5 \\
32 & 0.34381(6)  & 1.51(3) & & 32 & 0.34390(9) & 0.6 \\
48 & 0.34376(11) & 1.48(6) & & 48 & 0.34397(15) & 0.8 \\
64 & 0.34399(11) & 1.61(7) & & 64 & 0.34378(22) & 0.1 \\
96 & 0.34385(15) & 1.6(1)  & & & \\
128& 0.34387(24) & 1.5(2)  & & & \\
\end{tabular}
\end{ruledtabular}
\end{table}

\begin{table}
\caption{Estimates of $T_c$ and $\lambda$ obtained 
  by minimizing the functions (\ref{chi2def})
  (superscript 1) and (\ref{chi2def2}) (superscript 2),
  varying the minimal value $L_{\rm min}$ of the size allowed in the
  fits.  
  We also report $s^{(i)} \equiv A^{(i)}_{\rm min}/(N_L-N_f)$, where
  $A_{\rm min}$ is the minimum of the function $A^{(i)}$,
  cf. Eqs.~(\ref{chi2def}), (\ref{chi2def2}), $N_L$ and $N_f$ are
  the number of $L$-values used and the number of
  fit parameters, respectively. }
\label{resfitsY}
\begin{ruledtabular}
\begin{tabular}{rllllll}
$L_{\rm min}$ & $T_c^{(1)}$ & $\lambda^{(1)}$ &  $s^{(1)}$ &
$T_c^{(2)}$ & $\lambda^{(2)}$ &  $s^{(2)}$ \\
\colrule
4  & 0.34374(3) & 1.484(5) & 8.8 & 0.34387(2) & 1.532(7)  & 1.5 \\  
6  & 0.34380(1) & 1.502(2) & 0.6 & 0.34382(2) & 1.510(6)  & 0.5  \\  
8  & 0.34381(1) & 1.504(2) & 0.5 & 0.34382(2) & 1.508(8)  & 0.6  \\  
12  & 0.34381(1) & 1.504(4) & 0.6 & 0.34385(4) & 1.527(14)  & 0.6  \\  
16 & 0.34382(2) & 1.509(6)  & 0.5 & 0.34385(4) & 1.53(2)  & 0.6  \\  
32 & 0.34384(3) & 1.517(14)  & 0.7 & 0.34393(8) & 1.59(6) & 0.7  \\  
64 & 0.34389(6) & 1.55(4)  & 0.9 & 0.34370(25)& 1.4(2)    & 1.0 \\  
96 & 0.34373(21) & 1.42(17)  & 1.6 & & & \\  
128 & 0.34374(15) & 1.41(12) & 0.8 & & & \\  
\end{tabular}
\end{ruledtabular}
\end{table}

We estimate $T_c$ in the BH model, by using the matching method
discussed above.  The most accurate results are obtained by analyzing
the helicity modulus $\Upsilon$.  Fig.~\ref{rawY} shows the QMC
estimates of $\Upsilon$ versus $T$ up to $L=256$. Note that they do
not show any particular feature which may hint at a critical
point. For instance, $T_c$ cannot be estimated by using the
crossing-point method, as is usually done in FSS analyses of standard
phase transitions, e.g., in the three-dimensional BH
model.\cite{CTV-13}

This is due to a peculiar feature of the behavior of RG invariant
quantities at the BKT transition.  Usually, the large-$L$ limit of
$R(L,T)$ is discontinuous at $T_c$, with i.e., three different results
are obtained in the limit $L\to \infty$, depending whether $T<T_c$,
$T=T_c$, $T> T_c$, i.e.
\begin{eqnarray}
&&\lim_{T\to T_c^\pm} \lim_{L\to\infty} R(L,T) \neq \lim_{L\to\infty} R(L,T_c). 
\label{difflim}
\end{eqnarray}
As a consequence, for finite values of $L$, this discontinuity appears
as a crossing of the finite-size curves around $T_c$.  In the case of
the 2D XY universality class, instead, the infinite-volume limit shows
no discontinuity as $T$ approaches $T_c$ from the low-temperature
side, i.e.
\begin{eqnarray}
\lim_{T\to T_c^-} \lim_{L\to\infty} R(L,T) =
\lim_{L\to\infty} R(L,T_c). 
\label{difflimXY}
\end{eqnarray}
Hence, a crossing of the finite-size curves does not necessarily
occur (it may still be present if scaling corrections have different
signs on the two sides of the transition, but this does not occur for
$\Upsilon$).

We implement the matching methods using various procedures, to
crosscheck the results.  The simplest one considers the data for only
two lattice sizes at once, $L$ and $2L$. Then, $\lambda(L)$ and an
effective critical temperature $\widetilde{T}(L)$ are determined by
solving the two equations
\begin{eqnarray}
&& \Upsilon[L,\widetilde{T}(L)] = \widetilde{\Upsilon}_{\rm XY}(\lambda(L) L), 
 \nonumber \\
&& \Upsilon[2 L,\widetilde{T}(L)] = \widetilde{\Upsilon}_{\rm XY}(2 \lambda(L) L).
\label{twopm}
\end{eqnarray}
To implement this strategy, we need accurate estimates of $\Upsilon$
in a sufficiently wide temperature interval close to $T_c$ for each
value of $L$. For this purpose, we performed QMC simulations at
relatively close values of $T$ and computed the first and second
derivatives of $\Upsilon$ with respect to $T$, obtaining accurate
interpolations close to the BKT transition.  The results of the the
two-point matching procedure based on Eqs.~(\ref{twopm}) are reported
in Table \ref{restwopY}.  They appear quite stable, indicating that
the residual power-law scaling corrections are small.  Indeed, the
solutions $\widetilde{T}(L)$ of Eqs.~(\ref{twopm}) are expected to
approach $T_c$ with corrections of order $L^{-2}$.  To take them into
account, we fit $\widetilde{T}(L)$ to $T_c + c/L^2$, including only
data satisfying $L\ge L_{\rm min}$. Results are stable, essentially
independent of $L_{\rm min}$, confirming that the neglected scaling
corrections are irrelevant within our typical error bars.

Instead of using only two lattice sizes at each step of the matching
procedure, one may adopt a more general strategy in which all results
satisfying $L\ge L_{\rm min}$ are used at once. In this alternative
approach, $T_c$ and $\lambda$ are determined by minimizing the
$\chi^2$-like function
\begin{eqnarray}
A^{(1)}(T,\lambda) = \sum_i \left[ {\Upsilon(L_i,T) -
    \widetilde{\Upsilon}_{\rm XY}(\lambda L_i) \over \Delta
    \Upsilon(L_i,T)}\right]^2,
\label{chi2def}
\end{eqnarray}
where the sum is over the available lattice sizes, and
$\Upsilon(L_i,T)$ and $\Delta \Upsilon(L_i,T)$ are the intepolating
curves with their errors.  In order to take into account the expected
$O(L^{-2})$ corrections, we also consider
\begin{eqnarray}
A^{(2)}(T,\lambda,c) = \sum_i \left[ {\Upsilon(L_i,T) - 
\widetilde{\Upsilon}_{\rm XY}(\lambda L_i) - c/L_i^2 
\over \Delta \Upsilon(L_i,T)}\right]^2 .
\label{chi2def2}
\end{eqnarray}
In Table~\ref{resfitsY} we report the results.  Indicative errors on
the optimal parameters $T_c$ and $\lambda$ are obtained from the
covariance matrix at the minimum of the functions (\ref{chi2def}) and
(\ref{chi2def2}), although this does not take into account all
statistical correlations of the quantities involved in the matching
procedure.  The role of the residual scaling corrections is checked by
increasing the minimal value $L_{\rm min}$ of the sizes considered in
the fit, and by comparing the results obtained by using
(\ref{chi2def}) and (\ref{chi2def2}).  It is clear that they are
negligible.

\begin{figure}[tbp]
\includegraphics[width=8.5cm, keepaspectratio=true]{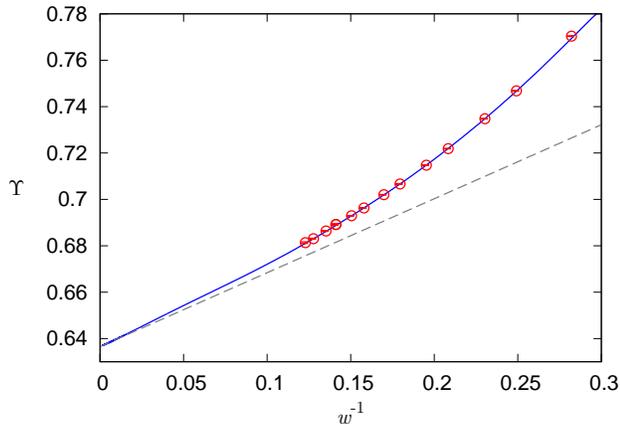}
\caption{(Color online) Plot of the BH estimates (circles) 
  of $\Upsilon$ at $T=0.3438\approx T_c$, 
  of the finite-size curve $\widetilde{\Upsilon}_{\rm XY}(L)$ (full line),
  and of the asymptotic prediction (\ref{rasym}) with constants (\ref{retalo})
  (dashed line).
  We take $\lambda = 1.5$ and set [see Eq.~(\ref{def-upsilon})]
  $w \equiv \ln(L/\Lambda) + {1\over 2}\ln\ln(L/\Lambda)$ 
  with $\Lambda_{\rm XY}=0.3$ for the XY model and 
  $\Lambda_{BH}=\Lambda_{\rm XY}/\lambda = 0.2$ for the BH model. 
  }
\label{Ymatching}
\end{figure}

Overall, the results of the analyses are quite stable. They indicate
\begin{equation}
0.3438 \lesssim T_c(\mu=0) \lesssim  0.3439,
\label{tcest}
\end{equation}
and $1.50\lesssim \lambda \lesssim 1.55$.  Fig.~\ref{Ymatching} shows
the quality of the matching procedure.  We report the BH estimates of
$\Upsilon$ at $T=0.3438\approx T_c$ and the critical XY curve
$\widetilde{\Upsilon}_{\rm XY}(\lambda L)$, assuming $\lambda = 1.5$.
The BH data fall on top of the XY curve, indicating that the
nonuniversal corrections of order $L^{-2}$ are quite small. In
Fig.~\ref{Ymatching} we also report the asymptotic prediction
(\ref{rasym}) with constants (\ref{retalo}). A comparison with the XY
curve $\widetilde{\Upsilon}_{XY}(L)$ shows that it describes the
finite-size behavior of $\Upsilon$ only for $w^{-1} \lesssim 0.05$,
hence for lattice sizes which are much larger than those we simulated,
up to $L=256$. Therefore, logarithmic fits to Eq.~(\ref{rasym}),
i.e. including only the leading logarithmic correction, would not
provide an accurate estimate for $T_c$.

\begin{figure}[tbp]
\includegraphics[width=8.5cm, keepaspectratio=true]{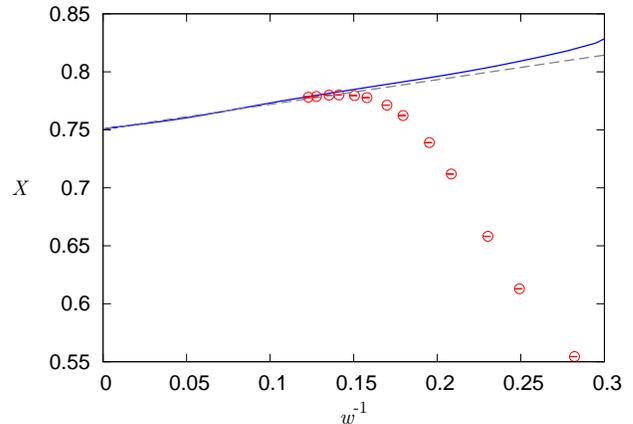}
\includegraphics[width=8.5cm, keepaspectratio=true]{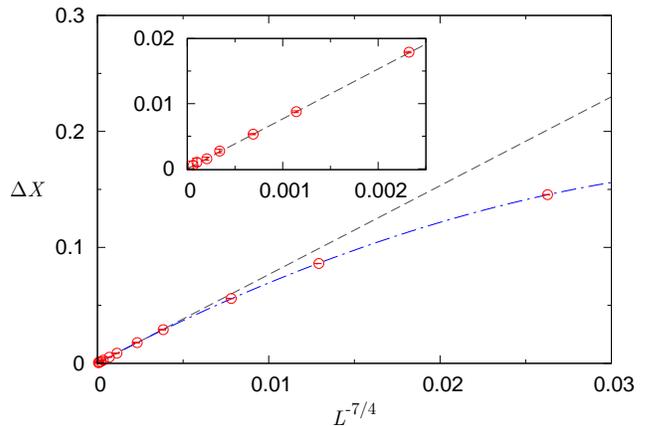}
 \caption{(Color online) Top: Plot of the BH estimates (circles) of $X\equiv
   \xi/L$ at $T=0.3438\approx T_c$, of the XY function $\widetilde{X}_{\rm XY}(L)$
   (full line), and of the asymptotic prediction (\ref{rasym}) with 
  constants (\ref{rxilo}) (dashed line). We use $\lambda=1.5$
  and set [see Eq.~(\ref{def-upsilon})]
  $w \equiv \ln(L/\Lambda) + {1\over 2}\ln\ln(L/\Lambda)$ 
  with $\Lambda_{\rm XY}=0.3$ for the XY model and
  $\Lambda_{BH}=\Lambda_{\rm XY}/\lambda = 0.2$ for the BH model. 
   Bottom: Difference
   $\Delta X= \widetilde{X}_{\rm XY}(\lambda L) - X(L,T_c)$ for $\lambda =
   1.5$.
   The dashed line is a linear fit to
   $aL^{-7/4}$ of the data satisfying $L\ge 24$. The dotted-dashed line shows a
   fit to $aL^{-7/4}+bL^{-2}+ cL^{-4}$ of the data for $L\ge 8$, which takes also
   into account the expected higher-order power-law corrections.
    }
 \label{rximatching}
\end{figure}

The estimate (\ref{tcest}) is confirmed by the data of $X\equiv
\xi/L$. At $T_c$ we expect the asymptotic behavior
\begin{equation}
X(L,T_c) = \widetilde{X}_{\rm XY}(\lambda L) + c L^{-7/4}, \label{xtcsca}
\end{equation}
which includes the expected $O(L^{-2+\eta})$ leading scaling
corrections.  The curve $\widetilde{X}_{\rm XY}(L)$ is reported in
App.~\ref{xycurves}.  Note that the rescaling $\lambda$ must be the
same as that obtained from the matching of $\Upsilon$, thus
$\lambda\approx 1.5$. In Fig.~\ref{rximatching} we compare the QMC
estimates of $X$ at $T=0.3438$ in the BH model with the XY curve
$\widetilde{X}_{\rm XY}(\lambda L)$ with $\lambda = 1.5$.  Note that
no adjustable parameters enter this comparison.  Here scaling
corrections are quite larger than those observed for the helicity
modulus.  Nonetheless, it is quite evident that Eq.~(\ref{xtcsca})
holds asymptotically, and that deviations scale as $L^{-7/4}$, as
predicted by theory.  In the figure, we also compare the BH data with
the asymptotic expression (\ref{rasym}) [coefficients are given in
  Eq.~(\ref{rxilo})].  In this case, the asymptotic prediction is
close to the curve $\widetilde{X}_{\rm XY}(L)$, indicating that, once
the power-law scaling corrections become negligible, a logarithmic fit
would provide the correct estimate of $T_c$.

\subsubsection{A more general analysis} \label{sec24.3}

\begin{table}
\caption{ Estimates of $T_c$ and $\lambda$ obtained 
  by fitting the data to Eq.~(\ref{exp-Ups-T-fit}) (superscript 1) 
  and to Eq.~(\ref{fit-withder2}) (superscript 2),
  varying the minimal value $L_{\rm min}$ of the size allowed in the
  analyses.  
  }
\label{resfitwithder}
\begin{ruledtabular}
\begin{tabular}{rllll}
$L_{\rm min}$ & $T_c^{(1)}$ & $\lambda^{(1)}$ & 
$T_c^{(2)}$ & $\lambda^{(2)}$   \\
\colrule
6   & 0.34384(2) &  1.52(1)  & 0.34384(2) & 1.52(1) \\
8   & 0.34385(2) &  1.52(1)  & 0.34384(2) & 1.52(1) \\
12  & 0.34386(3) &  1.53(2)  & 0.34385(3) & 1.53(2) \\
16  & 0.34386(3) &  1.53(2)  & 0.34386(3) & 1.53(2) \\
24  & 0.34386(3) &  1.54(2)  & 0.34386(5) & 1.53(3) \\
\end{tabular}
\end{ruledtabular}
\end{table}

The two analyses presented in Sec.~\ref{sec24.2} are straightforward
implementations of the matching procedure. However, they require
interpolations of the data at fixed size and an estimation of the
error on the interpolations.  One can avoid both steps, by slightly
generalizing the matching procedure.  For the values of $T$ that we
consider and for each size $L$, the helicity modulus is an almost
linear function of $T$, as is clear from Fig.~\ref{rawY}.  Therefore,
we can expand $\Upsilon(L,T)$ as
\begin{equation}
\Upsilon(L,T) \approx \Upsilon(L,T_c) + 
  (T - T_c) \left. {\partial \Upsilon \over \partial T}\right|_{T=T_c}.
\label{exp-Ups-T}
\end{equation}
Now, we use the matching condition to relate $\Upsilon(L,T_c)$ to the 
corresponding XY quantity. Including the leading scaling corrections
we obtain
\begin{equation}
\Upsilon(L,T) \approx \widetilde{\Upsilon}_{\rm XY} (\lambda L) + 
  (T - T_c) \left. {\partial \Upsilon \over \partial T}\right|_{T=T_c} + 
  c/L^2.
\label{exp-Ups-T-fit}
\end{equation}
This relation is at the basis of two different analyses of the BH
data, that differ in the treatment of the derivative term. 

In the first analysis we use the QMC results for $\partial
\Upsilon/\partial T$. Within errors, the estimates of the derivative
appear to be essentially independent of $T$ for $T$ close to
0.3438. Hence, to obtain a reliable interpolation we simply consider
all data in the interval $0.343\le T \le 0.345$ and determine an
interpolation of the form
\begin{eqnarray}
S(L) \equiv \left. {\partial \Upsilon \over \partial T}\right|_{T=T_c} = 
   w \sum_{k=0}^n \alpha_k w^{-k} + {c_T\over L^2},
\label{expansion-derivativeY}
\end{eqnarray}
where $w$ is defined in Eq.~(\ref{def-upsilon}) and we take $\Lambda =
\Lambda_{\rm XY}/\lambda = 0.2$ (we take $\Lambda_{\rm XY} = 0.3$,
consistently with the estimates reported in App.~\ref{xycurves}). The
interpolating function is consistent with the RG results of
App.~\ref{AppA}, which predict $\partial \Upsilon/\partial T \sim \ln
L$ at the critical point. We try several values of $n$, obtaining an
almost perfect interpolation for $n\ge 3$.  Once the interpolation is
known, we determine $T_c$ and $\lambda$ by fitting all BH data to
Eq.~(\ref{exp-Ups-T-fit}), taking $\lambda$, $T_c$, and $c$ as free
parameters. The fit is repeated several times, including each time
only the data satisfying $L\ge L_{\rm min}$, for an increasing
sequence of $L_{\rm min}$. The results are reported in
Table~\ref{resfitwithder}. We only take into account the statistical
errors and, in particular, we neglect the uncertainty on the
interpolation of the derivative of $\Upsilon$. Therefore, errors may
be (slightly) underestimated. The results are perfectly consistent
with those reported in Sec.~\ref{sec24.2} and confirm
Eq.~(\ref{tcest}).

We also performed a second fit, in which the QMC results for the
derivative of $\Upsilon$ are not used.  We only assume that an
expression like (\ref{expansion-derivativeY}) provides an accurate
interpolation. Therefore, we performed fits to
\begin{eqnarray}
&& \Upsilon(L,T) = 
 \widetilde{\Upsilon}_{\rm XY} (\lambda L) + {c \over L^2} + 
\nonumber \\
  && \quad (T - T_c) \left ( \alpha_0 h(L,\lambda) + \alpha_1 + 
    {\alpha_2\over h(L,\lambda)} + 
    {c_T\over L^2} \right),
\label{fit-withder2}
\end{eqnarray}
with 
\begin{eqnarray}
h(L,\lambda) = \ln {\lambda L\over \Lambda_{\rm XY}} + 
   {1\over2} \ln\ln {\lambda L\over \Lambda_{\rm XY}},
\end{eqnarray}
and $\Lambda_{\rm XY} = 0.3$. In the fit, $\lambda$, $T_c$,
$\alpha_0$, $\alpha_1$, $\alpha_2$, $c$, and $c_T$ are free
parameters. The results are reported in Table~\ref{resfitwithder}. In
spite of the quite large number of parameters, results are stable and
completely consistent with those obtained by using the interpolation
of the temperature derivative.  Again, they confirm Eq.~(\ref{tcest}).

\subsubsection{Final estimate of the critical temperature}
\label{sec24.4}

In the analyses presented in Sections~\ref{sec24.2} and \ref{sec24.3}
we made the implicit assumption that the exact curve for the helicity
modulus at criticality is known. But this is not the case.  There are two
sources of uncertainty. First, the interpolations are affected by an
error related to the statistical error of the XY data.  As discussed
in App.~\ref{xycurves}, this error is quite small.  The relative
uncertainty is less than $3\times 10^{-5}$ for $L\lesssim 400$, which
is the relevant region for the analysis of the BH data. This source of
error is practically irrelevant, since it changes the results of the
fits by a small fraction of the statistical error.  For the analyses
presented in Sec.~\ref{sec24.3}, $T_c$ varies by at most $4\times
10^{-6}$, if we change the curve by one error bar.

On the other hand, the error on the critical XY temperature gives rise
to systematic deviations on our final results which are not negligible
and which are of the order of the statistical errors.  To estimate the
corresponding systematic error, let us indicate with $\beta_{\rm XY} =
1.1199$ ($T_{\rm XY} = 1/\beta_{\rm XY}$), which is the value at which we
computed the XY function $\widetilde{\Upsilon}_{\rm XY}(L)$, and with
$T_{c,\rm XY}$ the {\em true} XY critical temperature.  If $\sigma =
T_{c,\rm XY} - T_{\rm XY}$ is small, we can write
\begin{eqnarray}
\Upsilon_{\rm XY}(T_{c,\rm XY},L) &\approx& 
    \Upsilon_{\rm XY}(T_{\rm XY},L) + \sigma S_{\rm XY}(L) 
\nonumber \\
   &=&
    \widetilde{\Upsilon}_{\rm XY}(L) + \sigma S_{\rm XY}(L),
\label{YtrueTc}
\end{eqnarray}
where $S_{\rm XY}(L) = \partial \Upsilon_{\rm XY}/\partial T$ 
computed at the critical point. For the BH model 
we can write analogously
\begin{equation}
\Upsilon_{\rm BH}(T,L) = \Upsilon_{BH}(T_{c,{\rm BH}},L) + (T-T_{c,{\rm BH}}) S_{\rm BH}(L).
\end{equation}
We can now use the matching conditions (\ref{matching-R}) and 
(\ref{matching-S}) to rewrite the previous relation as 
\begin{equation}
\Upsilon_{\rm BH}(T,L) = \Upsilon_{\rm XY}(T_{c,{\rm XY}},\lambda L) + 
    {1\over k}(T-T_{c,{\rm BH}}) S_{\rm XY}(\lambda L).
\end{equation}
Finally, Eq.~(\ref{YtrueTc}) implies
\begin{equation}
\Upsilon_{BH}(T,L) = \widetilde{\Upsilon}_{\rm XY}(\lambda L) + 
    \left[{1\over k}(T-T_{c,{\rm  BH}}) +\sigma\right] S_{\rm XY}(\lambda L).
\end{equation}
In our approach, we compute the value of $T$ which provides the best matching
between $\Upsilon_{\rm BH}(T,L)$ and $\widetilde{\Upsilon}_{\rm XY}(\lambda L)$. 
It corresponds to the value of $T$ for which the correction term vanishes,
hence 
\begin{equation}
   T = T_{c,{\rm BH}} - k \sigma .
\end{equation}
Therefore, the systematic error on our estimates of the critical
temperature of the BH model is $|k \sigma|$, where $\sigma$ is the
error on $T_{c,\rm XY}$: $\beta_{\rm XY}^2 \sigma = 10^{-4}$.  To
compute the nonuniversal constant $k$, we perform runs for the 
XY model at $\beta =
1.1201$ and a few values of $L$, obtaining estimates of $S_{\rm
  XY}(L)$.  Comparing this results with the BH ones, we estimate
$k/\beta_{\rm XY}^2 = 0.6(1)$, which gives 
\begin{equation}
k \sigma\lesssim 7\cdot 10^{-5}.  
\label{berr}
\end{equation}
If we now assume that the statistical error and the
systematic error due to the uncertainty on $T_{c,\rm XY}$ are
independent, we obtain the final estimate
\begin{equation}
T_c = 0.34385(9).
\label{Tc-BH-final}
\end{equation}

To conclude, we compare our result (\ref{Tc-BH-final}) with the
estimates of $T_c$ reported in the literature for the 2D hard-core BH
model at $\mu=0$: $T_c=0.3425(5)$ is reported in
Ref.~\onlinecite{CR-12}, $T_c=0.3423(3)$ in Ref.~\onlinecite{HK-97},
$T_c=0.353(3)$ in Ref.~\onlinecite{Ding-92}, and $T_c=0.350(4)$ in
Ref.~\onlinecite{DM-90}. There are significant discrepancies with our
result (\ref{Tc-BH-final}); we believe that this is essentially due to
the residual logarithmic corrections affecting the extrapolation of
the numerical data in Refs.~\onlinecite{CR-12,HK-97,Ding-92,DM-90}.
Other numerical results for generic 2D BH models can be found in
Refs.~\onlinecite{EM-99,CSPS-08,PKVT-08,RBRS-09,MDKKST-11,CT-12,YLFCW-12}.

\section{BKT critical behavior in a trap}
\label{trapan}

Experimental evidences of BKT transitions in trapped quasi-2D atomic
gases have been reported in
Refs.~\onlinecite{HKCBD-06,KHD-07,HKCRD-08,CRRHP-09,HZGC-10,Pl-etal-11,Desb-etal-12}.
The inhomogeneity due to the trapping potential drastically changes,
even qualitatively, the general features of the BKT critical behavior.
The analysis of the RG flow in the presence of a trapping potential
leads to nontrivial scaling laws characterized by the presence of
additional logarithmic factors.\cite{PV-13} Here, we discuss the
results of QMC simulations of the 2D hard-core BH model in the
presence of a harmonic trapping potential.  We verify the RG
predictions, and in particular the universality of the trap-size
dependence at the critical temperature.

\subsection{Trap-size dependence}
\label{tsstrap}

In sufficiently smooth inhomogeneous systems, the trap-size dependence
of the particle density $\rho({\bf x})\equiv \langle n_{\bf x}
\rangle$ can be approximately obtained by using the local-density
approximation (LDA).  In this approach, $\rho({\bf x})$ is
approximately given by the value of the particle density of the
homogeneous system at the same temperature and at an effective
chemical potential
\begin{equation}
\mu_{\rm eff}({\bf x}) = \mu - V(r) = \mu - (r/\ell_t)^2.
\label{mueff}
\end{equation}
Thus, the LDA implies that the particle density depends
on the ratio $r/\ell_t$ and that the total particle number asymptotically
scales as
\begin{equation}
N=\sum_{\bf x}\rho({\bf x})\approx c(\mu) \ell_t^2, 
\label{npa}
\end{equation}
with a coefficient that depends on $\mu$.  Actually, the LDA is expected
to give the exact space dependence of the particle density in the
large-$\ell_t$ limit keeping $r/\ell_t$ fixed. Scaling corrections 
decay as inverse powers of $\ell_t$,
\cite{CV-10-2,CTV-12} and are generally determined by
the universal features of the critical behavior.\cite{CV-10,CTV-13}

The main features of the critical behavior are encoded in the
correlation functions of the critical modes, like the one-particle
correlation function, whose leading behavior shows a nonanalytic TSS
around the critical point.  We consider the one-particle correlation
function between the center of the trap ${\bf x}=0$ and another point
${\bf y}$, i.e.
\begin{equation}
G_0({\bf y}) \equiv G({\bf 0},{\bf y}) 
\equiv \langle b_{\bf 0}^\dagger  b_{\bf y} \rangle.
\label{gbdef0}
\end{equation}
We also define the trap susceptibility
\begin{equation}
\chi_t = \sum_{\bf x} G_0({\bf x}), \label{defchitss}
\end{equation}
(it differs from the usual susceptibility $\chi$, which considers
correlations between any pair of points in the lattice), and the trap
correlation length $\xi_t$ defined in terms of the second moment of
$G_0({\bf x})$:
\begin{equation}
\xi_t^2 = {1\over 4 \chi_t} \sum_{\bf x} |{\bf x}|^2 G_0({\bf x}) .
\label{defxitss}
\end{equation}
At the BKT transition the TSS of $G_0({\bf x})$ presents
multiplicative logarithms,\cite{PV-13}
\begin{equation}
G_0({\bf x}) = \ell_t^{-1/4} (\ln \ell_t)^{1/8+\kappa/4}\,
 {\cal G}\left[{\bf x}(\ln \ell_t)^\kappa/\ell_t\right],
\label{g0xpc}
\end{equation}
where $\kappa$ is a new exponent related to the trap, which is
expected to depend on the main features of the external potential.  In
the inhomogeneous XY model (\ref{xymodtr}), $\kappa$ depends on the
power $q$ of the link function (\ref{hoppingt}). Numerical results
provide evidence of a simple dependence:~\cite{PV-13}
\begin{equation}
\kappa = 2/q.
\label{kq}
\end{equation}
Eq.~(\ref{g0xpc}) implies the scaling relations at $T_c$
\begin{eqnarray}
&&\chi_t  \sim  \ell_t^{7/4} (\ln \ell_t)^{1/8-7\kappa/4},\label{chitss}\\
&&\xi_t \sim  \ell_t (\ln \ell_t)^{-\kappa}.  \label{xitss} 
\end{eqnarray}
Note that the exponent $\kappa$ must vanish in the limit $q\to\infty$.
Indeed, for $q\to \infty$, one is dealing 
with a homogeneous system with open boundary conditions. Hence,
one should recover the usual FSS relations with $\ell_t \sim L$.

We would like to note that the additional logarithmic factors are a
specific feature of the BKT point. Indeed, in the low-temperature QLRO
phase the trap-size dependence is simpler. The correlation length
satisfies $\xi_t\sim \ell_t$ without logarithms,\cite{CV-11} and the
one-particle correlation function scales as
\begin{equation}
G_0({\bf x}) = \ell_t^{-\eta(T)}  {\cal G}\left({\bf x}/\ell_t\right),
\label{g0xpcqlro}
\end{equation}
with $\eta(T)<1/4$.

\subsection{Universality of the trap-size scaling}
\label{tsstrapuniv}

A major physical question concerns the universality of the trap
effects, i.e. under which conditions critical exponents and scaling
functions are the same for different trapped systems that belong to
the same ``homogeneous" universality class.

Let us begin by discussing which features of the trapping potential
are relevant in the TSS limit.  Consider a generic model with trapping
potential $V(r/\ell_t)$ where $r$ is the distance from the center of
the trap.  The TSS limit is obtained by taking $r\to\infty$,
$\ell_t\to\infty$ at fixed $\zeta = r/\ell_t^\theta$, where $\theta$
is the trap exponent.  Therefore, since $r/\ell_t = \zeta
\ell^{\theta-1}_t$, if $\theta < 1$ as generically occurs, in the TSS
limit the argument of the potential tends to zero: only the small-$r$
behavior, i.e., the one for $r\ll\ell_t$, is relevant.  For the XY
model we have $\theta = 1$ with logarithmic corrections controlled by
the exponent $\kappa$.\cite{PV-13} In this case, $\zeta = r (\ln
\ell_t)^\kappa/\ell_t$, so that $r/\ell_t =
\zeta/(\ln\ell_t)^\kappa$. Therefore, provided that $\kappa > 0$, also
in this case the relevant behavior is the small-distance one. As a
consequence, if $V(r/\ell_t) \approx (r/\ell_t)^p$ for $r\to 0$, in
all cases we expect TSS to depend only on $p$ and not on more specific
features of the potential.

In Ref.~\onlinecite{PV-13} scaling functions and exponent $\kappa$
were computed for the XY model (\ref{xymodtr}) with link function
$W(r)$ given in Eq.~(\ref{hoppingt}). They depend on the exponent $q$.
In particular, results were consistent with $\kappa = 2/q$. Since
$\kappa > 0$, they confirm that only the small-distance behavior of
$W(r)$ is relevant in the TSS limit. Since the BH model and the XY
model belong to same universality class in homogeneous conditions, the
trap-size behavior in the two models is expected to be strictly
related.

Since the small-$r$ behavior is the relevant property that
characterizes the universality class in the presence of the trap, it
seems natural to assume that the trapped BH model and the classical
inhomogeneous XY model have the same TSS at $T_c$ when $p=q$, i.e.,
when the trapping potential $V(r)$ and the link variable $W_{ij}(r)$
have the same short-distance behavior.  However, the scaling argument
presented below, which is confirmed by our numerical results, shows
that this is not always true. In particular, we argue that, at
$\mu=0$, the correct correspondence is $q=2p$.

For homogeneous systems, in the absence of a magnetic field, the
relevant quantity that controls the critical behavior of the 2D XY
model is the temperature difference $\tau = T - T_c$.  In the
inhomogeneous case, cf. Eq.~(\ref{xymodtr}), we can associate an
effective temperature $T_{\rm eff}(r)$ to each point at distance $r$
from the center of the trap.  Then, we argue that two inhomogeneous
systems have the same TSS provided that the short-distance behavior of
the relevant RG perturbation $\tau(r) = T_{\rm eff}(r) - T_c$ is the
same.  In the case of the classical XY model at $T_c$, we have
$\tau(r)\sim W(r)\sim r^q$.  In the case of the hard-core BH model at
$T=T_c(\mu)$, we have
\begin{equation}
\tau(r) \equiv T_c(\mu)-T_c[\mu_{\rm eff}(r)],
\label{taueff}
\end{equation}
where the effective chemical potential is given in Eq.~(\ref{mueff}).
Now, for $\mu \not = 0$ and $\delta\mu\ll 1$,
\begin{equation}
T_c(\mu + \delta\mu) - T_c(\mu) \sim \delta\mu.
\label{tcdeltamu}
\end{equation}
Therefore, at least for values of $r$ for which $V(r)\ll \mu$ (this is
the relevant region in the TSS limit, as discussed above), we obtain
$\tau(r) \sim V(r)\sim r^p$.  Therefore, we expect the same TSS for $p
= q$. This result, however, does not apply to the case $\mu = 0$,
due to the symmetry properties of the phase diagram reported in
Fig.~\ref{phd}. Because of the symmetry under $\mu\to -\mu$, we have
for $\mu\to 0$
\begin{equation}
T_c(\mu) = T_c(0) + c \mu^2 + O(\mu^4).
\label{tcmu}
\end{equation}
Then, by using again Eq.~(\ref{mueff}), we obtain $\tau(r) \sim
V(r)^2\sim r^{2p}$, which implies that, for $\mu = 0$ the correct
correspondence is $q = 2 p$.  This argument allows us to predict that,
for the BH model with $p=2$, the exponent $\kappa$ should be equal to
1 for $\mu\not=0$ and to 1/2 for $\mu = 0$.

The equality of the short-distance behavior of the effective $\tau(r)$
appears to be a necessary requirement for two systems to belong to the
same TSS universality class.  If this condition holds, critical
exponents are the same.  For the scaling functions however, a more
careful analysis is needed.  Indeed, in homogeneous systems, the
critical behavior depends on the phase: critical exponents are always
the same, but scaling functions in the high- and low-temperature phase
differ. Hence, when comparing scaling functions, one must be careful
to perform the comparison for systems that are in the same effective
phase.  It is therefore important to understand the effective behavior
in the inhomogeneous XY and BH models.

In the trapped hard-core BH model with $\mu\le 0$ and the XY model
(\ref{xymodtr}), the external field brings the system toward the
high-temperature phase when moving out of the center of the trap.  In
the XY model (\ref{xymodtr}) the effective space-dependent temperature
defined in Eq.~(\ref{teffxy}) tends to infinity when $r\equiv |{\bf
  x}|\to\infty$; in particular, for $T=T_c$, $T_{\rm eff}({\bf x}) >
T_c$ for any $r>0$.  Analogously, in the BH model one may define the
effective space-dependent chemical potential (\ref{mueff}).  When
$\mu\le 0$ and $T=T_c$, the phase diagram (\ref{phd}) is such that the
system is effectively in the high-temperature phase for any
$r>0$. Therefore, the hard-core BH model with $\mu\le 0$ and in the
inhomogenous XY model (\ref{xymodtr}) should share the same universal
features.  The two model should share the same TSS scaling functions,
provided the exponents satisfy the relations discussed above.

The behavior for $\mu > 0$ is expected to be different.  Since the
trapping potential decreases the local effective chemical potential
(\ref{mueff}), the system is in the superfluid QLRO phase for $0 <
r\lesssim \sqrt{2\mu} \ell_t$, and in the normal phase only for
$r\gtrsim \sqrt{2\mu} \ell_t$.  In the TSS limit at $\zeta = r (\ln
\ell_t)^\kappa/\ell_t$ fixed, we expect a low-temperature behavior for
$\zeta\lesssim \sqrt{2\mu} (\ln \ell_t)^\kappa$. Since $\kappa > 0$,
as $\ell_t$ increases, the boundary of the QLRO phase goes to
infinity: In the TSS limit all points belong to the QLRO phase.
Therefore, even if the TSS logarithmic exponent is the same as in the
case $\mu < 0$, scaling functions for $\mu > 0$ should differ from
those obtained for $\mu < 0$. Actually, we guess that the TSS
functions at $\mu>0$ should correspond to those of the inhomogenous XY
model (\ref{xymodtr}) with $U_{{\bf x}{\bf y}}=[1 + W(r_{{\bf x}{\bf
      y}})]^{-1}$ replaced by $U_{{\bf x}{\bf y}}= 1 + W(r_{{\bf
    x}{\bf y}})$.

An analogous scenario is also expected in the three-dimensional (3D)
hard-core BH model, whose phase diagram is similar to that reported in
Fig.~\ref{phd}, with a finite-temperature XY superfluid transition
line from ($\mu=-3,T=0$) to ($\mu=3,T=0$).  Again this transition line
is symmetric with respect to $\mu\to -\mu$, therefore $T_c(\mu)$ has a
maximum at $\mu=0$.  The TSS at the superfluid transition is generally
characterized by the trap exponent~\cite{CV-09} $\theta=
p\nu/(1+p\nu)$ for a trapping potential $V(r)=(r/\ell_t)^p$, where
$\nu=0.6717(1)$ is the correlation-length exponent of the 3D XY
universality class.\cite{CHPV-06} This is confirmed by a numerical
study~\cite{CTV-13} of the 3D BH model with a harmonic trapping
potential at $\mu=-2$, for which $\theta=0.57327(4)$.  This result
is expected to hold for all values of $\mu$ between $-3$ and $3$ that are
different from zero.  At the particular value $\mu=0$, the argument
outlined above implies that the trap exponent is different, being
associated with an external field given by $V(r)^2=(r/\ell_t)^{2p}$.
Therefore,\cite{CV-09} we expect $\theta=2p\nu/(1+2p\nu)$, giving
$\theta=0.72876(3)$ for $p=2$.  Finally, note that, although we expect
$\theta=0.57327(4)$ for any $\mu\not=0$, scaling functions should
depend on the sign of $\mu$.

\subsection{QMC results}
\label{qmctrap}

\begin{figure}[tbp]
\includegraphics*[scale=\graphicscale]{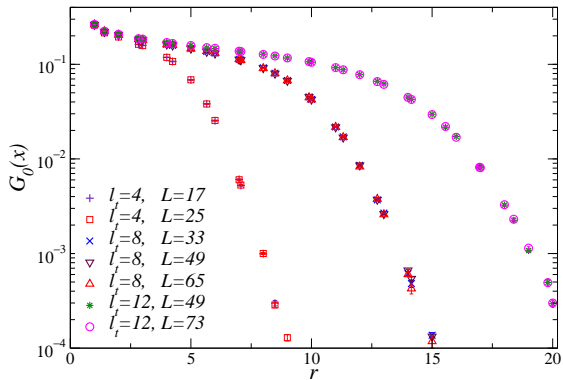}
\caption{(Color online) Comparison of 
  $G_0({\bf x})$ versus $r=|{\bf x}|$ for several 
   trap sizes $\ell_t$ and lattice sizes $L$, at $T=0.3438\approx T_c$.
   No size dependence is observed for $L\gtrsim 4\ell_t$. }
\label{ggftss}
\end{figure}

\begin{figure}[tbp]
\includegraphics*[scale=\graphicscale]{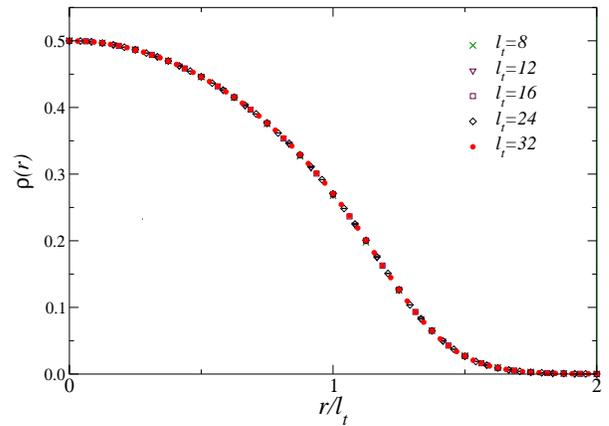}
\caption{(Color online) The particle density at $T=0.3438\approx T_c$ for
  several trap sizes versus $r/\ell_t$
(data obtained taking $L\gtrsim 4\ell_t$).  Data collapse
  onto a unique curve. }
\label{trapde}
\end{figure}

\begin{figure}[tbp]
\includegraphics*[scale=\graphicscale]{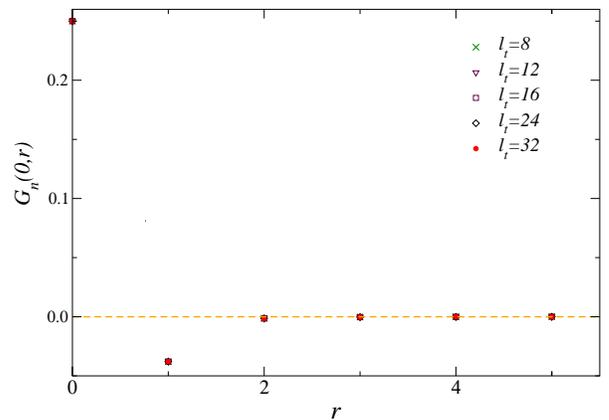}
\caption{(Color online) The connected density-density correlation
  $G_n(0,{\bf x})$ at
  $T=0.3438\approx T_c$ for several trap sizes, versus $r\equiv |{\bf x}|$. 
}
\label{trapdeco}
\end{figure}

We numerically check the RG predictions of the previous section by
performing QMC simulations of the 2D hard-core BH model (\ref{bhmt})
at $\mu=0$ and $T=0.3438\approx T_c$, with a harmonic potential [$p=2$
  in Eq.~(\ref{potential})]. We present results for several values of
the trap size $\ell_t$, up to $\ell_t=32$.  The trap is centered in
the middle of a square $L^2$ lattice, with odd $L$ and open boundary
conditions. More details on our practical implementation of QMC
simulations of trapped systems can be found in
Refs.~\onlinecite{CTV-12,CT-12}.

The lattice size $L$ is large enough ($L\approx 4 \ell_t$) to make
finite-size effects negligible compared with the statistical errors,
at least for the critical correlations with respect to the center of
the trap.  This is checked by comparing results for several lattice
sizes $L$ at fixed trap size $\ell_t$.  In Fig.~\ref{ggftss} we show
the one-particle correlation function $G_0({\bf x})$.  Within the
precision of our data, results obtained by using lattice sizes $L = 4
\ell_t + 1$ --- i.e. $L/\ell_t\approx 4$ --- can be
identified as infinite-volume results.

\begin{figure}[tbp]
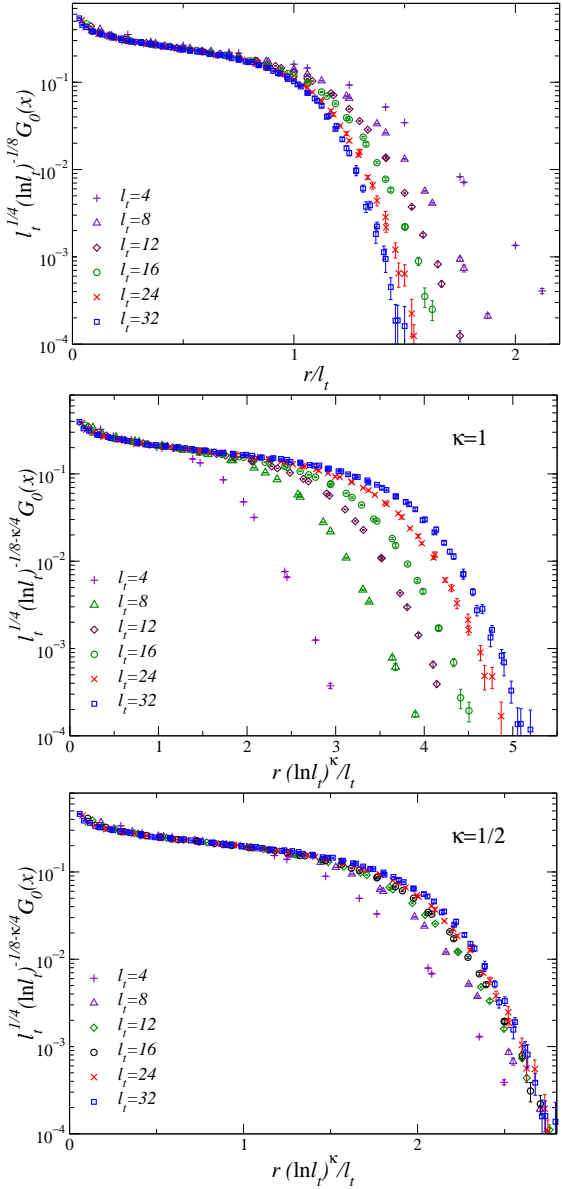

\includegraphics*[scale=\graphicscale]{fig9a.eps}
\includegraphics*[scale=\graphicscale]{fig9b.eps}
\includegraphics*[scale=\graphicscale]{fig9c.eps}
\caption{(Color online) Scaling of the one-particle correlation
  function $G_0({\bf x})$ in a trap, using Eq.~(\ref{g0xpc}) with
  $\kappa=0$ (top, note that this  is analogous to the rescaling of the
  distance for the particle density, as in Fig.~\ref{trapde}),
  $\kappa=1$ (middle) and $\kappa=1/2$ (bottom). The value
  $\kappa=1/2$ is clearly favoured by the data, supporting the RG
  arguments of Sec.~\ref{tsstrapuniv}.  All results are obtained for
  $L=4\ell_t+1$.~\cite{footnoteftss} }
\label{ggtrap}
\end{figure}

In Fig.~\ref{trapde} we show the particle density $\rho({\bf x})\equiv
\langle n_{\bf x} \rangle$ at $T_c$. The data collapse onto a single
curve when plotted as a function of the rescaled distance
$r/\ell_t$. The large-$\ell_t$ convergence at fixed ratio $r/\ell_t$
is quite fast; one can hardly see variations for $\ell_t\ge 8$
already.  These results are consistent with the LDA, as discussed in
Sec.~\ref{tsstrap}. In particular, we find $\rho(0) \approx 1/2$ (we
obtain $\rho(0)=0.5006(3),\;0.4999(3),\;0.5004(4)$ respectively for
$\ell_t=8,\,16,\,32$), which is the particle density of the
homogeneous system at $\mu=0$, cf. Eq.~(\ref{rhoh}).  The data in
Fig.~\ref{trapde} show also that, at $T=0.3438$ and $\mu=0$, the
particles are effectively confined within a spherical region $|{\bf
  x}|\lesssim 2\ell_t$ by the harmonic potential (indeed $\rho \approx
10^{-4}$ for $r\approx 2\ell_t$).

The connected
density-density correlation
\begin{equation}
G_n({\bf 0},{\bf x})\equiv 
\langle n_{\bf 0} n_{\bf x}\rangle-
      \langle n_{\bf 0}\rangle \langle n_{\bf x}\rangle
\label{grhodef}
\end{equation} 
vanishes rapidly, see Fig.~\ref{trapdeco}, without showing any
particular scaling behavior. This is analogous to the behavior
observed in the one-dimensional BH model in the zero-temperature
quantum superfluid phase.\cite{CTV-12} Notice that the value $G_n(0,0)
\approx 1/4$ at the center of trap (we obtain $G_n(0,0)=1/4$ for any
$\ell_t$ with a precision of $10^{-6}$) agrees with the corresponding
LDA prediction.  Indeed, it corresponds to the value $G_n({\bf x},{\bf
  x})=1/4$ in homogeneous systems at zero chemical
potential.\cite{footnotegrho}

\begin{figure}[tbp]
\includegraphics*[scale=\graphicscale]{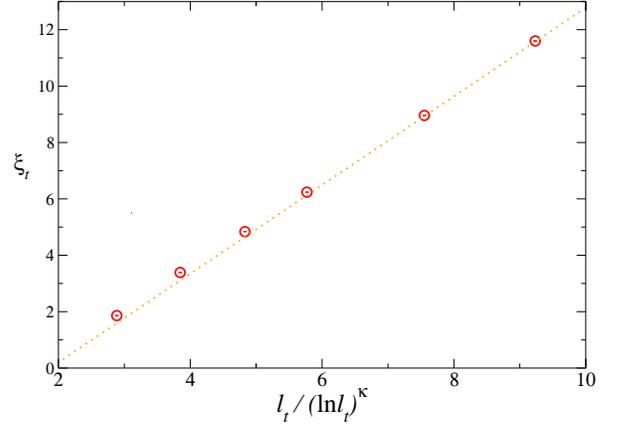}
\caption{(Color online) QMC data of the second-moment correlation
  length at $T=0.3438\approx T_c$ versus
  $\ell_t(\ln \ell_t)^{-\kappa}$, with $\kappa = 1/2$. 
   The dotted straight line is drawn to guide the eye.  }
\label{xit}
\end{figure}

Fig.~\ref{ggtrap} shows QMC data for the one-particle correlation
function $G_0({\bf x})$.  They are rescaled using Eq.~(\ref{g0xpc}),
with $\kappa=0,\,1/2$, and 1.  The best collapse of the data is
observed for $\kappa=1/2$, supporting the RG arguments of the previous
section.  In Fig.~\ref{xit} we show the trap correlation length
$\xi_t$. The data are consistent with the asymptotic behavior with
$\kappa = 1/2$, i.e., with
\begin{equation}
\xi_t  \sim \ell_t (\ln \ell_t)^{-1/2}.
\label{xitlt1o2}
\end{equation}

\begin{figure}[tbp]
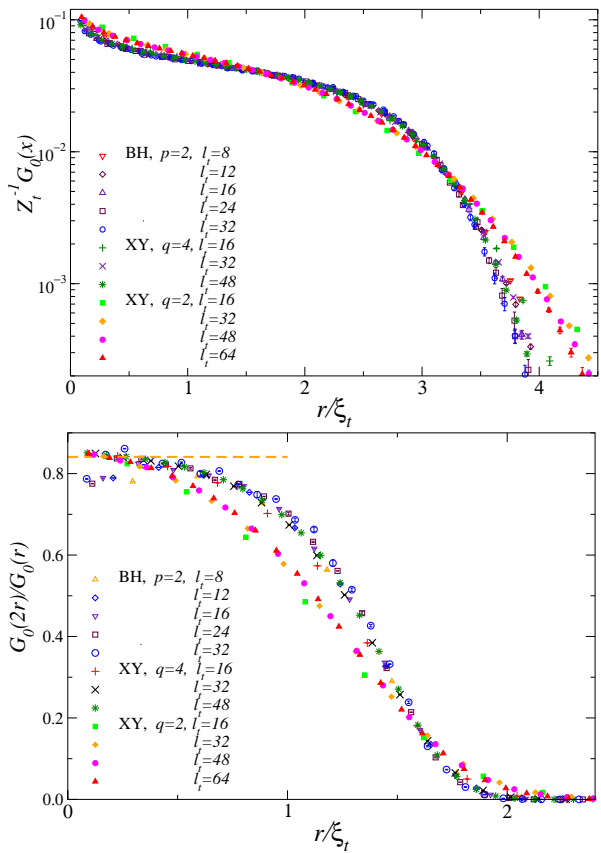

\includegraphics*[scale=\graphicscale]{fig11a.eps}
\includegraphics*[scale=\graphicscale]{fig11b.eps}
\caption{(Color online) Universality check at $T_c$:
  plot of $Z_t^{-1}
  G_0(x)$ with $Z_t = \chi_t/\xi_t^2$ (top) and of the ratio $G_0(2
  r)/G_0(r)$ (bottom) vs $r/\xi_t$ with $r\equiv |x|$, for the trapped
  hard-core BH model with $p=2$, and for the classical XY model with $q=4$
  and $q=2$.  In both cases the data for the harmonically
  ($p=2$) trapped BH model converge to the 
  large-$\ell_t$ curve of the classical XY model with a quartic
  external field ($q=4$). In the bottom panel the horizontal dashed
  line gives the $r\to\ 0$ limit (in the continuum limit).}
\label{ggtrapuniv}
\end{figure}

We now perform a more stringent universality check, verifying that the
hard-core BH model with $p=2$ and the classical XY model at $T_{\rm
  XY}$ with $q=4$ have the same TSS. To avoid the use of free
parameters, we plot the correlation functions in terms of the
universal ratio $r/\xi_t$ instead of $r/\ell_t$, and consider
$G_0({\bf x})/Z_t$, $Z_t \equiv \chi_t/\xi_t^2$, and $G_0(2 {\bf
  x})/G_0({\bf x})$. With these choices, the nonuniversal
normalization of the correlation function cancels out.  In
Fig.~\ref{ggtrapuniv} we compare the BH results for these two
quantities with the corresponding data for the classical XY model
(\ref{xymodtr}) with $q=2$ and $q=4$. The XY data obtained by using
$q=4$ fall on top of the BH results, while those obtained by using
$q=2$ differ significantly. Hence, these results provide a nice
evidence for the correspondence $q = 2p$, as argued in the previous
section.

\section{Conclusions}
\label{conclusions}

We investigate the universal behavior of quantum 2D many-body systems
at finite-temperature BKT transitions.  We consider the 2D hard-core
BH model (\ref{bhm}) [which is equivalent to the so-called XX spin
  model (\ref{XX})] in homogeneous conditions and in the presence of a
harmonic trap, cf. Eq.~(\ref{bhmt}).  The phase diagram, sketched in
Fig.~\ref{phd}, presents a BKT transition line between the normal and
superfluid QLRO phase.  We consider the system at half filling ($\mu =
0$), performing extensive QMC simulations. We first perform a detailed
FSS analysis to determine the critical temperature and then a detailed
check of the TSS RG predictions.

In order to determine the critical BKT temperature, we employ a
numerical matching method, which relates the FSS of RG invariant
quantities in the 2D BH model with that in the classical XY model
(\ref{XYmodel}) at its critical temperature, which is known with high
accuracy.\cite{HP-97,Hasenbusch-05} In this approach, the residual
scaling corrections decay as inverse powers of the lattice size $L$
and not as powers of $1/\ln L$, as in more standard approaches.
Therefore, it allows us to perform reliable large-$L$ extrapolations
and to obtain accurate and robust estimates of the critical
temperature.  This procedure yields $T_c(\mu=0)=0.34385(9)$, which
significantly improves earlier
estimates.\cite{CR-12,HK-97,Ding-92,DM-90}

We stress that the matching method can be used to determine the
critical temperature of any physical system undergoing a BKT
transition, using the XY finite-size curves reported in
App.~\ref{xycurves}.  The precision of the method is essentially
limited by the relative precision of the estimate $\beta_{\rm
  XY}=1.1199(1)$ of the XY critical temperature. This limitation may
be overcome by computing the relevant FSS curves in the BCSOS model,
for which $T_c$ is exactly known.\cite{Baxter-82,HMP-94}

The matching method is quite general and can be used in other systems
characterized by marginal RG perturbations, as long as an accurate
estimate of the critical temperature for at least one representative
of the given universality class is known.  For instance, one could
apply it to the study of three-dimensional tricritical transitions, of
$\Phi^4$ scalar models in four dimensions,\cite{It-Dr-book} of the
quantum $T=0$ Mott transition of the 2D BH model,\cite{FWGF-89,SSS-94}
or of two-dimensional randomly dilute Ising
systems.\cite{RandomlyDilute,HPPV-08}

We also investigate the universal BKT behavior in the presence of a
harmonic trap, cf. Eq.~(\ref{bhmt}).  In this case the trap exponent
takes the trivial value $\theta=1$, but, additionally, logarithmic
corrections appear. \cite{PV-13} At the critical BKT temperature the
correlation length $\xi_t$ scales as $\xi_t \sim \ell_t (\ln
\ell_t)^{-\kappa}$ with respect to the trap size $\ell_t$, where the
exponent $\kappa$ depends on the general features of the external
trapping potential. For the classical XY model with an effective
space-dependent temperature $T_{\rm eff}(T,{\bf
  x})=T[1+(r/\ell_t)^q]$, numerical simulations \cite{PV-13} suggested
the relation $\kappa = 2/q$.  We argue, and provide numerical
evidence, that the trapped BH models and this inhomogeneous XY model
share universal features.  The exponent $\kappa$ should generally be
equal to $2/p$ [$p$ is defined in Eq.~(\ref{potential})], hence we
should have $\kappa = 1$ for a harmonic potential.  This result does
not apply to the particular case of the hard-core BH model at zero
chemical potential.  We argue that in this case the BH model with
potential exponent $p$ is in the universality class of the XY model
with $q = 2p$. Hence $\kappa = 1/2$ for $p = 2$. Note that, although
the exponent $\kappa$ is the same for any $\mu\not=0$, scaling
functions are expected to depend on the sign of $\mu$.

To verify the theoretical arguments, we perform extensive QMC
simulations of the hard-core BH model at $\mu=0$.  We consider several
trap sizes, up to $\ell_t=32$, measuring the one-particle correlation
function $G_0({\bf x})=\langle b_0^\dagger b_{\bf x}\rangle$ between
the center of the trap ${\bf 0}$ and a generic point ${\bf x}$.  A
careful analysis of the numerical results, presented in
Sec.~\ref{qmctrap}, confirms the scaling arguments, in particular the
correspondence between the BH model with $p=2$ with the XY model with
$q=4$.
 
The results we present in this paper are relevant for experiments
probing BKT transitions, both in homogeneous systems, such as $^4$He
at the superfluid transition,\cite{GKMD-08} and in inhomogeneous
systems, such as cold atoms in harmonic traps.\cite{BDZ-08} In
particular, experimental results for the one-particle correlation
function of quasi-2D trapped atomic gases can be inferred from the
interferences between two atomic clouds.\cite{HKCBD-06,CRRHP-09} The
analysis of experimental data reported in Ref.~\onlinecite{HKCBD-06}
provided some evidence of an algebraic decay in the superfluid
regime~\cite{HKCBD-06}, although trap effects were effectively
neglected.

Accurate studies of the critical properties of trapped systems, to
check universality and determine the critical exponents, require a
robust control of the effects of the confining potential.  Our results
should be useful to determine the critical parameters of the BKT
transition, without requiring further assumptions and approximations,
such as mean-field and local-density approximations, to handle the
inhomogeneity arising from the trapping force. In particular, in
trapped quasi-2D particle systems one may exploit the difference of
the trap-size dependence of the critical correlation functions between
the high- and low-temperature phases: with decreasing $T$ we go from
the normal phase, in which the trap-size dependence is trivial, to the
QLRO phase where the length scale increases proportionally to the trap
size, through the BKT transition where the relation between $\xi_t$
and $\ell_t$ shows also multiplicative logarithms, $\xi_t \sim \ell_t
(\ln \ell_t)^{-\kappa}$.

\acknowledgements We thank Martin Hasenbusch for useful
correspondence, in particular for providing us Monte Carlo data for
the classical XY model at $T_{\rm XY}$.  The QMC simulations were
performed at the INFN Pisa GRID DATA center, using also the cluster
CSN4.

\appendix

\section{Off-critical finite-size scaling: renormalization-group predictions}
\label{AppA}

In Ref.~\onlinecite{PV-13} we investigated the FSS behavior at the BKT
critical point.  We wish now to discuss the nature of the off-critical
corrections.  For a standard phase transition, these corrections can be
expressed in power series of $\tau L^{1/\nu}$, $\tau = (T -
T_c)/T_c$. Therefore, critical-point scaling holds as long as $\tau
L^{1/\nu} \ll 1$.  Since formally $\nu = \infty$, in the XY model we
expect some expansion in terms of $\tau \ln^\alpha L$. We wish now to
show that this guess is correct and moreover we wish to compute the
exponent $\alpha$. We find
\begin{equation}
\alpha = 2.
\end{equation}
We start from the flow of $v(l)$ with $l_0 = \ln L$ (notations are those
of  Ref.~\onlinecite{PV-13}):
\begin{equation}
\ln L = \int_{v(L)}^{v_0} 
   {dw\over [Q + F(w)] (1 + w f(w^2))} .
\end{equation}
If we now consider the leading term for $Q\to 0$ and $L\to \infty$, we 
obtain 
\begin{equation}
\ln L \approx \int_{v(L)}^{v_0} 
   {dw\over [Q + w^2]} =
    {1\over \sqrt{Q}} \left({\rm arctan}{v_0\over \sqrt{Q}} - 
    {\rm arctan}{v\over \sqrt{Q}}\right).
\end{equation}
It follows 
\begin{equation}
v(L) = {v_0 - \sqrt{Q} \tan (\sqrt{Q} \ln L) \over 
        1 + v_0 \tan (\sqrt{Q} \ln L)/\sqrt{Q} }.
\end{equation}
In order to obtain critical-point behavior, we should assume 
that $\sqrt{Q} \ln L\ll 1$. Since $Q\sim \tau$, this implies 
$\tau \ln^2 L \ll 1$, which proves the relation $\alpha = 2$.
If we now expand the previous relation in powers of $\sqrt{Q} \ln L$,
we obtain the first-order temperature correction:
\begin{equation}
v(L) = {v_0\over 1 + v_0 \ln L} 
     - {Q \over 3} {\ln L (v_0^2 \ln^2 L + 3 v_0 \ln L + 3) \over 
                    (1 + v_0 \ln L)^2}.
\end{equation}
If we now consider the infinite-volume limit, this gives
\begin{equation}
v(L) = {1\over \ln L}\left(1 - {Q \ln^2 L\over 3}\right) + \ldots
\end{equation}
Let us now consider a renormalization-group quantity $R$, which satisfies
\begin{equation}
R(Q,v_0,L) = R(Q,v(L),1).
\end{equation}
Expanding this scaling equation in powers of $Q$ and $v$, we find 
\begin{eqnarray}
R(Q,v_0,L) &=& R^* + a v(L) + b Q  \nonumber  \\
  &\approx&
    R^* + {a\over \ln L} \left(1 - {Q \ln^2 L\over 3}\right) + \ldots
\end{eqnarray}
Since $Q\sim (T-T_c)$, this relation implies 
\begin{equation}
   \left. {\partial R\over \partial T}\right|_{T_c} = 
   - {a\alpha_1 \over 3} \ln L
\end{equation}
for $L\to \infty$.

\section{Exact relations within the QLRO phase}
\label{gauuniv}

The critical behavior of systems with $U(1)$ symmetry in the 
QLRO low-temperature phase can be 
obtained by studying the Gaussian spin-wave theory
\begin{equation}
H = {\beta\over 2} \int d^2 x \, (\nabla \varphi)^2.
\label{s2dxy}
\end{equation}
The relevant two-point function is 
\begin{equation}
G({\bf x}_1,{\bf x}_2) = 
\langle e^{-i\varphi({\bf x}_1)} e^{i\varphi({\bf x}_2)}\rangle.
\label{gdefgau}
\end{equation}
A simple calculation shows that, in the infinite-volume limit,  
$G({\bf x}_1,{\bf x}_2)\sim |{\bf x}_1-{\bf x}_2|^{-\eta}$, where the 
exponent $\eta$ is related to the coupling $\beta$ by
\begin{equation}
\eta = {1\over 2\pi\beta}.
\label{etade}
\end{equation}
This spin-wave theory describes the QLRO phase of 2D U(1)-symmetric
models for $\beta\ge 2/\pi$, corresponding to $0\le \eta \le 1/4$.
The values $\beta=2/\pi$ and $\eta=1/4$ correspond to the BKT
transition.

Let us now consider the system on a finite $L\times L$ square with
periodic boundary conditions.  The two-point function for the Gaussian
theory with periodic boundary conditions can be exactly computed,
obtaining \cite{CFT-book,It-Dr-book}
\begin{eqnarray}
G_{\rm sw}({\bf x}) = \left[
{e^{\pi z_2^2} \theta'_1(0,e^{-\pi})\over |
\theta_1[\pi(z_1+iz_2),e^{-\pi}]|}\right]^\eta,
\label{gcpbc}
\end{eqnarray}
where $z_i\equiv x_i/L$, $\theta_1(u,q)$ and $\theta_1'(u,q)$ are
$\theta$ functions.\cite{Gradstein} However, in the original model the
field $\varphi$ is periodic, hence the XY model with periodic boundary
conditions is equivalent to the Gaussian theory, provided one sums
over all twisted boundary conditions, i.e. over all configurations
satisfying $\varphi(L) = \varphi(0) + 2n\pi$, $n$ arbitrary integer.
This effect can be taken into account by the following
formula:\cite{Hasenbusch-05,CFT-book}
\begin{eqnarray}
&&G({\bf x}) = G_{\rm sw}({\bf x})\times\\
&&\quad {\sum_{n_1,n_2=-\infty}^{\infty} W(n_1,n_2) \cos[2\pi(n_1 x_1+n_2 x_2)]
\over \sum_{n_1,n_2=-\infty}^{\infty} W(n_1,n_2) },\nonumber \\
&&W(n_1,n_2) = \exp[-\pi(n_1^2+n_2^2)/\eta].
\nonumber
\end{eqnarray}
Using Eq.~(\ref{gcpbc}) and this expression, we can compute the universal 
function $X(\eta)$, where $X\equiv \xi/L$ and
\begin{eqnarray}
&&\xi^2 =  {L^2\over 4 \pi^2} \left( {\chi \over \chi_1}-1\right),\\
&&\chi =  \int d^2x \,G({\bf x}),\quad
\chi_1 =  \int d^2x \,{\rm cos}\left({2\pi x_1\over L}\right) 
\,G({\bf x}).
\nonumber
\end{eqnarray}
The curve $X(\eta)$ is shown in Fig.~\ref{gqlro}.  In particular,
close to the BKT point, i.e., for $\eta\to 1/4$, we obtain
\begin{eqnarray}
X(\eta) = 0.7506912222 + 1.699451\,\left( {1\over 4}-\eta\right) + \ldots
\label{rxitoro}
\end{eqnarray}
which is in agreement with the numerical results reported in
Ref.~\onlinecite{Hasenbusch-05}.  Analogous results have been obtained
for the helicity modulus,\cite{Hasenbusch-05}
\begin{eqnarray}
&&\Upsilon(\eta) = 
{1\over 2\pi\eta} - {\sum_{n=-\infty}^\infty n^2 \,{\rm exp}(-\pi n^2/\eta)\over
\eta^2 \sum_{n=-\infty}^\infty {\rm exp} (-\pi n^2/\eta)} =
\label{rupstoro}\\
&&= 0.6365081782 +2.551196 \, \left( {1\over 4}-\eta\right) + ...
\nonumber
\end{eqnarray}
The asymptotic behaviors (\ref{rasym}) for $X$ and $\Upsilon$ are obtained
by noting~\cite{Hasenbusch-05} that $1/4-\eta\approx 1/(8 w)$ for 
$w\to \infty$ and $\eta\to 1/4$.

\section{XY finite-size curves}
\label{xycurves}

We consider the classical XY model (\ref{XYmodel}) at 
$\beta_{\rm XY}\equiv 1/T_{\rm XY}=1.1199$ (which is the best 
available estimate\cite{HP-97} of the critical temperature), on $L\times L$
square lattices with periodic boundary conditions.  We determine 
accurate approximations of the functions $\widetilde{\Upsilon}_{\rm XY}(L)$
and $\widetilde{X}_{\rm XY} (L)$, by interpolating 
the available Monte Carlo data \cite{Hasenbusch-05,Hasenbusch-08} at
$\beta_{\rm XY}=1.1199$, which cover the interval $4\le L \le 4096$. 
Moreover, we also enforce the known large-$L$ asymptotic behaviors
(\ref{rasym}), (\ref{retalo}) and (\ref{rxilo}), to guarantee the correct 
limit $L\to \infty$. We consider functional forms motivated by theory. 
Hence, the functions $\widetilde{Y}_{\rm XY}(L)$ and $\widetilde{X}_{\rm XY}(L)$
are expressed as polynomials in the variable
$w \equiv \ln({L/\Lambda}) + {1\over2} \ln\ln(L/\Lambda)$,
 plus some correction terms of 
order $L^{-2k}$. In practice, we approximate the two functions as 
\begin{equation}
  \sum_{k=0}^n a_k/w^k + \sum_{k=1}^m b_k L^{-2k}.
\end{equation}
In the case of $\widetilde{X}$ we also add a term $c L^{-7/4}$, 
as suggested by the correction-to-scaling analysis presented in 
Sec.~\ref{sec22}.
Coefficients $a_0$ and $a_1$ are fixed by using Eqs.~(\ref{rasym}), 
(\ref{retalo}) and (\ref{rxilo}), while the others are obtained by fitting
the data. 

First, we performed several fits, changing the orders of the
polynomials $n$ and $m$, and considering the scale $\Lambda$ as a free 
parameter. We obtained almost perfect interpolations by taking 
$n = 5$ and $m = 3$. We also observed that the fits of $X$ and $\Upsilon$ 
were providing very similar values for the scale $\Lambda$. With the 
previous choices of $n$ and $m$, we obtained $\Lambda \approx 0.31$ from
the analysis of $\Upsilon$ and $\Lambda\approx 0.34$ from the analysis 
of $X$. Similar values where obtained in fits in which $n$ and $m$ were 
slightly changed. The consistency of the two results confirms the 
adequacy of our interpolating expressions. To obtain the final interpolations
we fixed $\Lambda = 0.3$ for both quantities, obtaining
\begin{eqnarray}
\widetilde{\Upsilon}_{\rm XY}(L) &=&  0.6365081782 
+ 0.318899454 \,w^{-1}   \label{eqytilde} \\         
&+& 2.0319176\, w^{-2}  - 40.492461 \, w^{-3}  
\nonumber\\
&+& 325.66533 \,w^{-4} - 874.77113 \,w^{-5} 
\nonumber \\
&+& 8.43794\,L^{-2} + 79.1227\,L^{-4} - 210.217\,L^{-6},  \nonumber
\end{eqnarray}
and
\begin{eqnarray}
\widetilde{X}_{\rm XY}(L) &=& 0.7506912222
+ 0.21243137 \,w^{-1}  \label{eqxtilde} \\         
&-& 2.8971137 \,w^{-2}  +76.328276 \, w^{-3}
\nonumber \\ 
&-& 642.23425 \,w^{-4} + 1805.3274 w^{-5} 
\nonumber \\
&-& 1.879629 \,L^{-7/4} - 15.00621 \,L^{-2} \nonumber \\
&-& 196.205 \,L^{-4} + 805.644 \,L^{-6}.  \nonumber
\end{eqnarray}
It is important to note that these expressions have the only purpose of 
providing an accurate interpolation of the numerical data, and are not 
meant to provide the correct large-$L$ asymptotic behavior. 
For instance, they should not be used to estimate the coefficient of the 
corrections proportional to $1/\ln^2 L\sim w^{-2}$: $a_2$ 
obtained in the fit probably differs significantly from the correct 
large-$L$ coefficient.

\begin{figure}[t]
\includegraphics*[scale=\graphicscale]{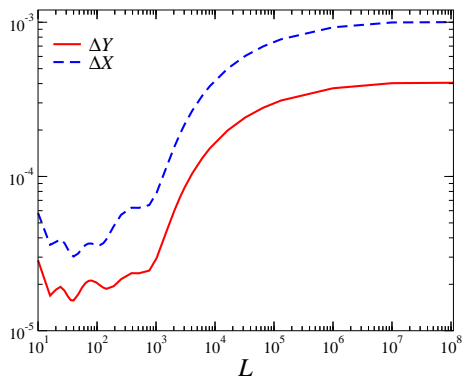}
\caption{(Color online) Plot of the errors on the curves
  (\ref{eqytilde}) and (\ref{eqxtilde}). 
  }
\label{errint}
\end{figure}

The uncertainty on the interpolation curves is related to the errors on the
available Monte Carlo data. It depends on $L$ and can be evaluated by using a
bootstrap method, which yields the curves shown in Fig.~\ref{errint}.
Errors are almost constant
and very small up to $L\approx 10^3$, i.e., in the region in which we have
data, then they increase, reaching a
maximum for $L \approx 10^8$. Eventually, they should  decrease to zero as
$1/(\ln L)^2$, due to the fact that the interpolations have the correct
asymptotic behavior~(\ref{rasym}).

\begin{thebibliography}{99}
\bibitem{mappingXX}
The spin operators are related to the boson operators $b_i$ by $\sigma^x_i =
b_i^\dagger + b_i$, $\sigma^y_i = i(b_i^\dagger - b_i)$, $\sigma^z_i =
1-2b_i^\dagger b_i$.
\bibitem{stats} Typical statistics of the runs range from $10^6$ to
  $5\times 10^7$ Monte Carlo steps, 
  where a Monte Carlo step is made of one diagonal
  update and $N_l$ off-diagonal updates.  The number $N_l$ is
  estimated at run time and turns out to linearly increase with $L$.
\bibitem{footnoteY} The helicity modulus $\Upsilon$ is related to the change
of the partition function $Z$ under twists $\phi$ of the boundary
 conditions along one of the lattice directions, i.e.  $\Upsilon = -
\partial^2 \ln Z(\phi)/\partial \phi^2|_{\phi=0}$.
\bibitem{footnoteftss} Since the correlation length scales as
  $\xi_t\sim \ell_t/(\ln \ell_t)^\kappa$, finite-size effects in the
  critical quantities (related to the ratio $L/\xi_t$) should further
  decrease keeping the ratio $L/\ell_t\approx 4$ constant while
  increasing $\ell_t$.
\end{thebibliography}
\end{document}